\documentclass[%
 aip,
 amsmath,amssymb,
 reprint,%
]{revtex4-1}

\usepackage{graphicx}
\usepackage{dcolumn}
\usepackage{bm}

\usepackage[utf8]{inputenc}
\usepackage[T1]{fontenc}
\usepackage{mathptmx}
\usepackage{etoolbox}

\usepackage[english]{babel}
\usepackage[utf8]{inputenc}
\usepackage[colorinlistoftodos, color=green!40, prependcaption]{todonotes}
\usepackage{amsthm}
\usepackage{mathtools}
\usepackage{physics}
\usepackage{xcolor}
\usepackage[left=23mm,right=13mm,top=35mm,columnsep=15pt]{geometry} 
\usepackage{adjustbox}
\usepackage{placeins}
\usepackage[T1]{fontenc}
\usepackage{lipsum}
\usepackage{csquotes}
\usepackage[pdftex, pdftitle={Article}, pdfauthor={Author}, colorlinks=true,linkcolor=blue,citecolor=blue,urlcolor=blue]{hyperref}
\usepackage{nameref}
\usepackage{changes}
\usepackage[babel]{microtype}

\usepackage{xr}
\makeatletter

\newcommand*{\addFileDependency}[1]{
\typeout{(#1)}
\@addtofilelist{#1}
\IfFileExists{#1}{}{\typeout{No file #1.}}
}\makeatother

\newcommand*{\myexternaldocument}[1]{%
\externaldocument{#1}%
\addFileDependency{#1.tex}%
\addFileDependency{#1.aux}%
}

\myexternaldocument{supporting_material}

\makeatletter
\def\@email#1#2{%
 \endgroup
 \patchcmd{\titleblock@produce}
  {\frontmatter@RRAPformat}
  {\frontmatter@RRAPformat{\produce@RRAP{*#1\href{mailto:#2}{#2}}}\frontmatter@RRAPformat}
  {}{}
}%
\makeatother

\newcommand{\rev}[1]{\textcolor{black}{#1}}

\newcommand{\A}{\mathrm{A}}
\newcommand{\C}{\mathrm{C}}
\newcommand{\U}{\mathrm{U}}
\newcommand{\G}{\mathrm{G}}
\newcommand{\ALD}{\langle\mathrm{ALD}\rangle}
\newcommand{\MLD}{\langle\mathrm{MLD}\rangle}
\newcommand{\Nbr}{\langle N_\mathrm{br}\rangle}
\newcommand{\nbr}{n_\mathrm{br}}
\newcommand{\Rg}{R_\mathrm{g}}
\newcommand{\Z}{\mathcal{Z}}
\newcommand{\F}{\mathcal{F}}
\renewcommand{\vec}{\bm}

\begin{document}

\preprint{AIP/123-QED}

\title[Scaling properties of RNA as a randomly branching polymer]{Scaling properties of RNA as a randomly branching polymer}

\author{Domen Vaupoti\v{c}}
\affiliation{Department of Theoretical Physics, Jo\v{z}ef Stefan Institute, Jamova 39, 1000 Ljubljana, Slovenia}
\author{Angelo Rosa}
\affiliation{Scuola Internazionale Superiore di Studi Avanzati (SISSA), Via Bonomea 265, 34136 Trieste, Italy}
\author{Luca Tubiana}
\affiliation{Department of Physics, University of Trento, via Sommarive 14, 38123 Trento, Italy}
\affiliation{INFN-TIFPA, Trento Institute for Fundamental Physics and Applications, via Sommarive 14, 38123 Trento, Italy}
\author{An\v{z}e Bo\v{z}i\v{c}}
\affiliation{Department of Theoretical Physics, Jo\v{z}ef Stefan Institute, Jamova 39, 1000 Ljubljana, Slovenia}
\email{anze.bozic@ijs.si}

\date{\today}

\begin{abstract}
Formation of base pairs between the nucleotides of an RNA sequence gives rise to a complex and often highly branched RNA structure. While numerous studies have demonstrated the functional importance of the high degree of RNA branching---for instance, for its spatial compactness or interaction with other biological macromolecules---
RNA branching topology remains largely unexplored. Here, we use the theory of randomly branching polymers to explore the scaling properties of RNAs by mapping their secondary structures onto planar tree graphs. Focusing on random RNA sequences of varying lengths, we determine the two scaling exponents related to their topology of branching. Our results indicate that ensembles of RNA secondary structures are characterized by annealed random branching and scale similarly to self-avoiding trees in three dimensions. We further show that the obtained scaling exponents are robust upon changes in nucleotide composition, tree topology, and folding energy parameters. Finally, in order to apply the theory of branching polymers to biological RNAs, whose length cannot be arbitrarily varied, we demonstrate how both scaling exponents can be obtained from the distributions of the related topological quantities of individual RNA molecules with fixed length. In this way, we establish a framework to study the branching properties of RNA and compare them to other known classes of branched polymers. By understanding the scaling properties of RNA related to its branching structure we aim to improve our understanding of the underlying principles and open up the possibility to design RNA sequences with desired topological properties.
\end{abstract}

\maketitle

\section{Introduction}

Branching structures can be found in a wide range of polymeric materials, including networks, gels, and, importantly, biopolymers such as nucleic acids and polysaccharides~\cite{Kulkarni2006,Voit2009,BiopolymerPhysics}. Compared to their linear counterparts, branching imparts polymers with several favourable properties such as high surface functionality, globular conformation, high solubilities, and more~\cite{Cook2020}. A ubiquitous, functionally important biopolymer is ribonucleic acid (RNA), which can be seen to effectively behave as a branched polymer due to the arrangement of single-stranded (ss) regions (such as loops, bulges, and stems) and double-stranded (ds) regions formed by base-pairing which can lead to the formation of multi-loops with a large degree of branching. Indeed, a survey of PDB-deposited RNA structures has shown that while multi-loops having branching degree 3 and 4 are the most abundant, branched loops with degree $\gtrsim 10$ are far from unusual~\cite{Wiedemann2022}. This propensity for high branching degrees sets RNA aside from most other known classes of branched polymers~\cite{Kulkarni2006,Voit2009}.

The branching structure of RNA also has a functional importance, which is the most apparent in the genomes of ssRNA viruses~\cite{Boerneke2019,Wan2022}. Recent studies on these viruses have shown that branching affects their assembly by both influencing the ability of RNA to bind to capsid proteins as well as making the size and the structure of the folded RNA genome comparable to the size of the capsid~\cite{Yoffe2008,Tubiana2015,Singaram2015,Garmann2016,Beren2017,Bozic2018,Marichal2021,Zandi2009,vds2013,Zandi2015}. Furthermore, the branching pattern combined with electrostatic interactions impacts the RNA osmotic pressure inside the capsid and in this way influences both its packaging efficiency and virion stability~\cite{Gonca2014,Erdemci2016}.
All of these properties are arguably the result of an evolutionary pressure. For instance, the \mbox{ssRNA} genomes of icosahedral viruses are significantly more compact than random RNA sequences of comparable length and composition~\cite{Yoffe2008,Gopal2014}. At the same time, even a small percent of synonymous mutations suffices to destroy this compactness and make the size of viral RNAs indistinguishable from that of random RNAs~\cite{Tubiana2015,Bozic2018}.

While some attempts have been made to experimentally determine the branching patterns of long RNAs from two-dimensional (2D) projections~\cite{Garmann2015}, for the most part such analysis still needs to be carried out using computational predictions of RNA secondary structure. In absence of pseudoknots, the secondary structure of RNA can be mapped to a planar tree graph by mapping its various structural elements (stems, bulges, hairpin loops, \ldots) to the tree vertices and edges~\cite{Gan2003,Schlick2018,Vaupotic2022}. Under this approximation, a basic measure of RNA size is its maximum ladder distance (MLD), which corresponds to the diameter of the tree and correlates well with its physical size as experimentally measured by, for instance, its hydrodynamic radius~\cite{Gopal2014,Borodavka2016}. Such an approach has been previously used to extensively analyze the compactness of viral ssRNA genomes~\cite{Yoffe2008,Gopal2014,Tubiana2015,Bozic2018}. What is more, mapping RNA structures to planar tree graphs also opens up the possibility to apply the powerful framework of polymer physics---and Flory theory in particular~\cite{FloryChemBook,Giacometti2013,Everaers2017}---to study the properties of long RNA molecules. \rev{We provide an overview of some of the theoretical approaches to study the branching topology of the secondary structures of viral ssRNA genomes and its implications in Ref.~\onlinecite{Vaupotic2022}.}

An ensemble of polymer conformations can be described in terms of a relatively small number of observables. Since the number of total accessible conformations of a polymer scales with its number of monomers (bonds) $N$, one can adopt the average value of an observable $\langle \mathcal O(N) \rangle$ to classify the polymer ensemble under consideration (such as ideal polymers, self-avoiding polymers, $\theta$-polymers, \ldots)~\cite{RubinsteinColbyBook}. Due to the scale-invariant character of polymer chains in the large-chain limit~\cite{RubinsteinColbyBook}, $\langle \mathcal O(N) \rangle$ behaves as a power law of the number of monomers $N$ of the chain with some characteristic scaling exponent $\gamma$:
\begin{equation}
\langle {\mathcal O}(N) \rangle \sim N^\gamma .
\end{equation}
In general, $\gamma$ depends on the physical properties of the monomers, for instance on how the monomers interact with each other or with the solvent~\cite{Wang2017}. Scaling exponents obtained from such power laws allow one to distinguish between branched polymers that live in 2D or 3D, or between polymers that are self-avoiding or at the $\theta$-point. They furthermore allow one to distinguish between different kinds of branching patterns---in particular, between regular patterns, such as observed in dendrimers, and random patterns, where the size and disposition of the branches are probabilistic~\cite{RubinsteinColbyBook}. In most cases, the aim of such a description is to obtain a scaling relationship for observables connected to the molecule size, as these can be experimentally measured. This was the goal of Fang {\em et al.}~\cite{Fang2011}, who used secondary structure prediction and Kramers' formula~\cite{Kramers1946,LubenskyIsaacson1979,DaoudJoanny1981} to obtain the scaling exponent $\nu$ for the radius of gyration of long RNAs. 

\rev{However, in the case of branching polymers, two more exponents, $\rho$ and $\varepsilon$, are required to get a more complete picture of their properties~\cite{vanRensburg1992,Everaers2017}. The first exponent, $\rho$, describes the scaling of the average shortest-path distance between two nodes on the tree. The second exponent, $\varepsilon$, describes the scaling of the average branch weight, namely, the average weight of the smaller of the two sub-trees obtained by removing the edges connecting two nodes of the original tree.} These properties are purely topological and their exponents are expected to capture the statistics of the branching structure. For example, when the topology of a random branching is annealed (i.e., when the branches can rearrange themselves), these two exponents coincide~\cite{vanRensburg1992}.
The scaling relation for the radius of gyration can then be obtained from these exponents and from the knowledge of the kind of solvent the polymer is immersed in. Despite the importance of the branching structure of RNA, its \rev{relationship} to other types of branched polymers remains unknown~\cite{Grosberg2018,Kelly2016}, and the scaling properties relating to the topology of branching are largely unexplored~\cite{Yoffe2008,Tubiana2015,Singaram2015}.

In this work, we set to determine the scaling exponents $\rho$ and $\varepsilon$ of RNA secondary structures \rev{obtained from random RNA sequences}, compare them with known types of branching polymers, and demonstrate how these exponents connect to the radius of gyration of ensembles of RNA folds using Flory theory. Our approach allows us to decouple the assumptions made to obtain the scaling of the radius of gyration from those related to the topology of the folding. We focus on \rev{\em arbitrary (random) RNA sequences}, which represent the baseline for RNA structure formation and also show the general behaviour one can expect of biological RNAs~\cite{Higgs1993,Bundschuh2002,Schultes2005,Clote2005,Chizzolini2019}. Using ViennaRNA software to predict the thermal ensembles of their secondary structures, we explore the importance of nucleotide composition, folding energy parameters, and node degree distribution on the scaling behaviour of RNA. Importantly, we also demonstrate how scaling exponents can be determined for \rev{random} RNA sequences with fixed length, which is particularly relevant for biological RNAs, whose length cannot be varied arbitrarily. Our results, which yield $\rho\simeq\varepsilon\approx0.67$, show that although ViennaRNA produces secondary structure folds only on the basis of energetic considerations, thus ignoring steric clashes and tertiary interactions, these folds scale as 3D self-avoiding annealed branched polymers, independently of the energy parameters. Interestingly, our results also show that, in a good solvent, the radius of gyration of RNA scales as $N^{1/2}$ and not as $N^{1/3}$ as estimated previously~\cite{Fang2011}.

The paper is organized as follows: In Sec.~\ref{sec:Theory} we provide an overview of the theory of branching polymers and introduce the topological observables applicable to RNA secondary structure, their corresponding distribution functions, and their scaling exponents. In Sec.~\ref{sec:Methods}, we describe the folding algorithm we use to predict RNA secondary structure and describe its tree representation. We also provide details on the important parameters, such as multiloop energy model and RNA sequence composition, that we make use of in our analysis. In Sec.~\ref{sec:Results}, we describe and compare the results for the scaling exponents of random RNA sequences obtained in two different ways. We furthermore verify the robustness of the obtained results by varying the multiloop parameters of the energy model, sequence composition, and topology of branching. The same section also frames RNA in terms of a randomly branching polymer and compares its scaling behaviour to that of other known types of branching polymers. Finally, in Sec.~\ref{sec:Discussion} we discuss our results in a wider context and show what the scaling exponents obtained from the topology of RNA structure tell us about the scaling of its size as given by the radius of gyration.

\section{Scaling theory of branching in RNA molecules}\label{sec:Theory}

\subsection{Average polymer behaviour}\label{ssec:avg-pol}
In this work, we build on a comparison between RNA secondary structure folds and randomly branched polymers. A large body of computational work~\cite{vanRensburg1992,RosaEveraersJPA2016,RosaEveraersJCP2016,RosaEveraersPRE2017,Everaers2017} has demonstrated that in order to completely characterize an ensemble of branched polymers, it is necessary to consider the topology of branching, also known as {\em tree connectivity}~\cite{Everaers2017}. This is a property that is accessible using simple RNA folding models (described in Sec.~\ref{sec:Methods}).

As a proper measure of tree connectivity in branched RNA molecules, we introduce the ensemble average of either the maximum (MLD) or average ladder distance (ALD) as a function of the number of monomers $N$~\cite{RosaEveraersJPA2016,Vaupotic2022},
\begin{equation}\label{eq:L-definition}
\langle\mathrm{MLD}(N)\rangle\sim\langle\mathrm{ALD}(N)\rangle \sim N^\rho , 
\end{equation}
both of which account for the average length of linear paths on the tree. Here, the ladder distance (path length) $\ell_{ij}$ between two nodes $i$ and $j$ is defined as the shortest path between them, and we thus have $\mathrm{MLD}=\max_{i,j}\ell_{ij}$ and $\mathrm{ALD}=[N(N+1)]^{-1}\sum_{i\neq j}\ell_{ij}$~\cite{Vaupotic2022}. \rev{Note that the average $\langle\mathcal{O}(N)\rangle$ is taken over an ensemble of trees of size $N$, which for random RNAs means an average over both sequences and secondary structures (see Sec.~\ref{ssec:ranRNA} and~\ref{ssec:RNAfol}).}

As another topological observable, we consider the average branch weight~\cite{vanRensburg1992}:
\begin{equation}\label{eq:Nbr-definition}
\langle N_{\rm br}(N) \rangle \sim N^\varepsilon ,
\end{equation}
which is defined as the average weight $n_\mathrm{br}$ of the smallest of the two sub-branches obtained by systematically removing---one at a time---edges connecting two nodes of the original tree, so that $N_\mathrm{br} =\overline{n}_\mathrm{br}$, where the bar indicates the average over all branch weights in a single fold~\cite{RosaEveraersJPA2016} (see also Sec.~\ref{ssec:tree-rep} and Fig.~\ref{fig:1}).

While the two scaling exponents $\rho$ and $\varepsilon$ describe very different quantities, they are not independent from each other. In fact, by making very minimal assumptions on the randomly branching architecture of the polymers, the relation
\begin{equation}\label{eq:Epsilon=Rho}
\rho = \varepsilon   
\end{equation}
is expected to hold in general for annealed branching polymers~\cite{vanRensburg1992}. Accurate numerical proofs of Eq.~\eqref{eq:Epsilon=Rho} are given in Refs.~\onlinecite{vanRensburg1992,RosaEveraersJPA2016} for isolated self-avoiding trees in spatial dimensions from $2$ to $9$ and in Ref.~\onlinecite{RosaEveraersJCP2016} for melts of trees in 2D and 3D.  Confirming the validity of Eq.~\eqref{eq:Epsilon=Rho} for RNA structures would be a strong argument in support of the hypothesis that RNA molecules can also be modeled as randomly branching polymers.

Even the simplest theory for the characterization of secondary structure of generic branching polymers (and RNA molecules in particular) thus has to operate with these two distinct observables, $\langle \mathrm{ALD}(N) \rangle$ and $\langle N_{\rm br}(N) \rangle$.
Furthermore, obtaining the scaling exponents $\rho$ and $\varepsilon$ related to these two observables allows us to obtain the scaling exponent $\nu$ for the radius of gyration as well. We elaborate on this connection in Sec.~\ref{sec:Discussion}, where we also explain in some detail the possible connection between RNA secondary and tertiary structure, in particular with regard to the mean radius of gyration of an ensemble of molecules.

\subsection{Distribution functions}
Beyond the averages of observables introduced in the previous section, another fundamental part of information on the scaling exponents $\rho$ and $\varepsilon$ is contained in their corresponding distribution functions~\cite{RosaEveraersPRE2017}. In particular, theoretical considerations show that the path length distribution functions $p(\ell)$ in trees of size $N$ collapse onto {\em universal} master curves when plotted as a function of the rescaled path length $x=\ell / \langle \mathrm{ALD}(N) \rangle$~\cite{RosaEveraersPRE2017}:
\begin{equation}\label{eq:pl}
p(\ell) = \frac1{\langle \mathrm{ALD}(N) \rangle}\  q \left(\frac{\ell}{\langle \mathrm{ALD}(N) \rangle}\right) .
\end{equation}
These master curves are described well by the one-dimensional Redner-des Cloizeaux (RdC) function~\cite{RosaEveraersPRE2017}:
\begin{equation}\label{eq:q_RdC_l}
q(x) = C \, x^{\theta}\, \exp \left( -(K x)^{t} \right) ,
\end{equation}
whose numerical constants~\cite{Theta-ell-WhyNotation}
\begin{eqnarray}
C & = & t \, \frac{\Gamma^{\theta+1}((\theta+2)/t)}{\Gamma^{\theta+2}((\theta+1)/t)} , \label{eq:RdC_C_l} \\
K & = & \frac{\Gamma((\theta+2)/t)}{\Gamma((\theta+1)/t)} , 
\label{eq:RdC_K_l}
\end{eqnarray}
follow from the conditions that $p(\ell)$ is normalized to $1$ and that its first moment, $\langle \mathrm{ALD}(N) \rangle$, is the only relevant scaling variable. Here, $\Gamma(x)$ denotes the gamma function. Importantly, the Pincus exponents $\theta$ and $t$ in Eq.~\eqref{eq:q_RdC_l} are both related to the scaling exponent $\rho$ as~\cite{RosaEveraersPRE2017}:
\begin{eqnarray}
\theta & = & \frac{1}{\rho}-1 , \label{eq:theta_l} \\
t & = & \frac{1}{1-\rho} . \label{eq:tl}
\end{eqnarray}

In a similar fashion, it can be shown~\cite{RosaEveraersPRE2017} that the probability distribution of branch weights $p(\nbr)$ in a randomly branching tree of total size $N$ is accurately described by the Kramers-like~\cite{Kramers1946,LubenskyIsaacson1979,DaoudJoanny1981} formula:
\begin{equation}\label{eq:KramersTheory}
p(\nbr) = \frac{\Z_{\nbr} \Z_{N-\nbr-1}}{\sum_{\nbr=0}^{N-1} \Z_{\nbr} \Z_{N-\nbr-1}} , 
\end{equation}
where $\Z_{\nbr}$ is the total number (i.e., the partition function) of branched polymers with $\nbr$ edges. Next, we can take into account that the denominator in Eq.~\eqref{eq:KramersTheory} is just $\Z_N$, which in general scales as $\Z_N \sim c^N / N^{2-\varepsilon}$ for branching polymers~\cite{LubenskyIsaacson1979,DaoudJoanny1981,RosaEveraersPRE2017}, where $c$ is a numerical prefactor related to the tree branching probability per node (or, equivalently, the fraction of the tree branching points). In this way, Eq.~\eqref{eq:KramersTheory} takes on the simple scaling form:
\begin{equation}\label{eq:pn}
\frac{p(\nbr)}{N^{2-\varepsilon}} \simeq c^{-1} \left( \, \nbr \, (N-\nbr-1) \, \right)^{-(2-\varepsilon)} .
\end{equation}
Expanding on the approach by~\citet{RosaEveraersPRE2017}, we tentatively set an ansatz for the partition function $\Z_{\nbr}$ to
\begin{equation}\label{eq:ansatz}
\Z_{\nbr} = \frac{I_\beta(2\lambda \nbr)}{\left(\lambda \nbr \right)^\beta} , 
\end{equation}
where $I_\beta$ is the first modified Bessel function of order $\beta=3/2-\varepsilon$. This allows us to fit the analytical distribution $p(\nbr)$ to the entire range $0 \leqslant \nbr \leqslant N/2$. The justification of the ansatz in Eq.~\eqref{eq:ansatz} is given in the supplementary material, where we demonstrate that it correctly describes the scaling behaviour of branched polymers as given by Eq.~\eqref{eq:pn}.

The {\em analytical} functions for the distributions of the path lengths [Eq.~\eqref{eq:pl}] and branch weights [Eq.~\eqref{eq:KramersTheory}] have the advantage that we can use them to determine the scaling exponents $\rho$ and $\varepsilon$, respectively, even for polymers whose length dependence of observables is not known or cannot be easily obtained---such as is the case of biological RNAs, whose length is often constrained to a narrow range within a particular type, function, or species. Due to the finiteness of the polymers, these distributions in general show finite-size effects which disappear in the long-chain limit, as we shall also see in Sec.~\ref{ssec:fixed-len}.

\section{Methods}
\label{sec:Methods}

\subsection{Random RNA sequences}
\label{ssec:ranRNA}
Nucleotide frequency $f(b)$, $b\in\{\A,\C,\G,\U\}$, is the simplest statistical property of the primary sequence of RNA, which can nonetheless have a decisive influence on its secondary and tertiary structure. We define a uniformly random RNA sequence as one where all nucleotides are represented equally, $f(b)=1/4\;\forall b$. Deviations from this uniform composition can be then evaluated by the Euclidean distance $\delta^2=\sum_{b\in\{\A,\C,\G,\U\}}(f(b)-1/4)^2$, which correlates well with improved statistical measures such as the Jensen-Shannon divergence.

Since the nucleotide composition of biological RNAs often differs from the uniform one, we generate not only uniformly random RNA sequences but also random RNA sequences with different nucleotide frequencies. Specifically, we choose $16$ different nucleotide compositions which cover the convex hull of the space of $\sim 1800$ viral genomes of different positive single-stranded RNA viruses (see Table~\ref{tab:nuccomp} in the supplementary material) and as such represent the most extreme cases within this dataset. Random RNA sequences are obtained by first generating a sequence with a desired nucleotide composition and length and then randomly shuffling it. In general, we consider random RNA sequences from $100$ to $13500$ nt in length, where we generate for each random RNA of a given length and composition $200$ different random shuffles.


%
\subsection{RNA folding}
\label{ssec:RNAfol}
When an RNA molecule folds, it often does not adopt a single, well-defined conformation~\cite{Woods2017}. Therefore, it is more accurate to discuss an ensemble of RNA conformations instead of a single structure. We generate thermal ensembles ($T=37\;{}^\circ\mathrm{C}$) of RNA secondary structures using the \texttt{RNAsubopt} routine from the ViennaRNA software (v2.4.14)~\cite{Lorenz2011} with default settings. The secondary structure prediction does not include pseudoknots, which is a typical simplification that allows us to study the structures of even very long RNAs ($\sim10^4$~nt). For each RNA sequence, we generate $500$ secondary structure folds, sampled from the thermal ensemble.

\subsection{Tree representation of RNA secondary structure}\label{ssec:tree-rep}
Each RNA secondary structure without pseudoknots can be mapped to a planar tree graph. The simplest way to construct such a tree is by mapping double-stranded regions (base pairs) to {\em edges} with weights $w$ corresponding to the stem lengths, while single-stranded regions (unpaired nucleotides) are mapped to {\em nodes} connecting the edges~\cite{Vaupotic2022}. (Note that this procedure differs slightly from the RNA-as-graph method~\cite{Schlick2018} in order to avoid disconnected graphs which are sometimes produced by it.) The resulting number of edges $N$ of the RNA tree corresponds to the number of bonds in a branched polymer.

\begin{figure}[!t]
  \centering
  \includegraphics[width=\linewidth]{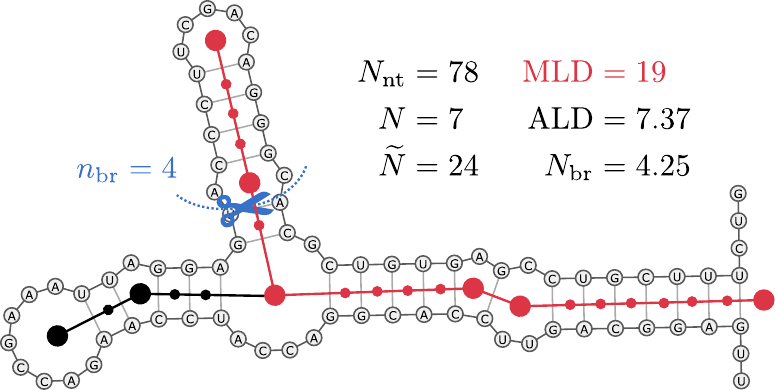}
  \caption{
  Tree representation of an example RNA secondary structure obtained for a random RNA sequence of length $N_\mathrm{nt}=78$. Single-stranded regions are mapped to nodes of a tree (large dots) while double-stranded regions are mapped to edges (lines), weighted with the length of the double-stranded region. Expanded tree produces additional nodes (small dots) between the nodes of the original tree corresponding to the weight of the edge, so that all the edges in the expanded tree have unit weight. Dotted line shows an example of removing one edge of the expanded tree to determine its branch weight $\nbr$, given by the weight of the smaller of the two resulting trees. The sketch also shows some essential properties of the resulting RNA tree, such as the number of edges in the original ($N$) and the expanded ($\widetilde{N}$) tree, and the MLD, ALD, and $N_\mathrm{br}$ of the tree.}
\label{fig:5}
\end{figure}

Furthermore, we produce what we term {\em expanded RNA trees}, where each edge of weight $w$ is expanded to form $w$ unit edges with $w-1$ nodes of degree $2$ connecting them (Fig.~\ref{fig:5}). In this way, we obtain a tree with $\widetilde{N}$ edges where each edge has unit weight, which allows us to use the percolation algorithm from Ref.~\onlinecite{RosaEveraersPRE2017} to obtain the branch weight distributions. \rev{The number of expanded edges $\widetilde N$ furthermore corresponds to the number of base pairs in the RNA secondary structure.} Since the RNA nucleotide sequence length $N_\mathrm{nt}$ perfectly correlates with both the number of edges and the number of expanded edges of the tree representation of its structure (Fig.~\ref{fig:S1} in the supplementary material), \rev{regardless of the nucleotide composition}, we use these quantities interchangeably throughout this work.

\subsection{Pr\"ufer-shuffled RNA trees}

\rev{Pr\"ufer sequence representation of RNA trees gives a one-to-one correspondence between a tree with $N$ edges and a sequence of $N-1$ integers. The Pr\"ufer sequence is generated iteratively from a labelled tree by successively removing the peripheral node with the smallest label, and adding the number of the node to which it was connected as the next element in the sequence. Random permutations of the Pr\"ufer sequence can then be used to yield trees with node degree distribution identical to the original tree but with different branching patterns. For a detailed description of the generation of the Pr\" ufer sequence of a tree and its permutations, see Ref.~\onlinecite{Singaram2016}.}

We utilize the Pr\"ufer sequence representation of RNA trees as an additional point of comparison to generate Pr\"ufer-shuffled versions of random RNA trees, which preserve the node degree distribution of the original trees. For each RNA length, we select $10$ different random sequences with $200$ secondary structures for each, and use the resulting trees to generate $500$ random permutations of their Pr\"ufer sequence.

\subsection{Multiloop energy model}
While the occurrence of multiloops of degree $10$ or higher is not uncommon in various RNAs~\cite{Wiedemann2022}, multiloop energies are the least accurately known among the numerous energy parameters involved in RNA secondary structure prediction~\cite{Poznanovic2021}. Most of the current energy-based structure prediction software---including ViennaRNA---assumes that the energy of a multiloop depends only on the amount of enclosed base pairs (number of branches) and the number of unpaired nucleotides in it, and uses a linear model of the form
\begin{eqnarray}
\label{eq:multiloop}
\nonumber E_\mathrm{multiloop}
& = & E_0 \nonumber\\
&  & + \ E_\mathrm{br} \times [\mathrm{branches}] \nonumber\\
&  & + \ E_\mathrm{un} \times [\textrm{unpaired nucleotides}] ,
\end{eqnarray}
where $E_0$ is the energy contribution for multiloop initiation, and $E_\mathrm{br}$ and $E_\mathrm{un}$ are the energy contributions for each enclosed base pair and unpaired nucleotide, respectively. 
Notable difference between different proposed energy parameters lies not only in the magnitude but in {\em the sign} of the parameter $E_\mathrm{br}$ which controls the number of branches stemming from the multiloop (see Ref.~\onlinecite{Vaupotic2022} for a comparison of multiloop energy parameters used in different energy models).
Earlier versions of ViennaRNA (until v2.0), for instance, used a positive value of this parameter, penalizing high-degree nodes, while the later versions of the software use a negative value, promoting high-degree nodes.

Differences in the multiloop energy parameters can of course reflect in the predicted structures of long RNAs and their topological properties~\cite{Ward2017,Poznanovic2020}.
This needs to be taken into account when comparing results obtained by existing studies on the branching properties of RNA~\cite{Gopal2014,Borodavka2016,Yoffe2008,Tubiana2015} that use different versions of folding software and thus potentially different energy models. To verify how the choice of multiloop energy parameters influences the scaling properties of RNA in the cleanest fashion, we modify {\em solely} the multiloop parameters in the current parameter set used by ViennaRNA (v2.4: $E_0=9.3$, $E_\mathrm{un}=0.0$ $E_\mathrm{br}=-0.9$) with the ones from older versions ($<$~v2.0: $E_0=3.4$, $E_\mathrm{un}=0.0$ $E_\mathrm{br}=0.4$), resulting in a modified set of energy parameters which we denote by ViennaRNA-mod. All other parameters are left unchanged, i.e., equal to the set used in ViennaRNA v2.4. We then compare the scaling of the RNA secondary structures obtained using the default parameter set (ViennaRNAv2.4) with those obtained using the modified set (ViennaRNA-mod).

\subsection{Scaling exponents}
We determine the two scaling exponents $\rho$ and $\varepsilon$ introduced in Sec.~\ref{sec:Theory} in two ways. Firstly, we determine the scaling of the ensemble-averaged topological properties (such as $\ALD$, $\MLD$, and $\Nbr$) with sequence length $N_\mathrm{nt}$.
Coefficients of the scaling power laws of the form $y=ax^\gamma$ can then be obtained with linear regression on logarithmically transformed data.  Where necessary to differentiate these exponents, we label them as $\rho_{\ALD}$, $\rho_{\MLD}$, and $\varepsilon_{\langle N_\mathrm{br}\rangle}$.

Secondly, we determine the scaling exponents $\rho$ and $\varepsilon$ also from the distributions of path lengths [Eq.~\eqref{eq:pl}] and branch weights [Eq.~\eqref{eq:KramersTheory}] in the tree representation of their secondary structure, respectively. Here, regression is inappropriate as it can in general give a biased estimate for the scaling exponents~\cite{Clauset2009}.
We therefore use the maximum likelihood estimator for the scaling exponents. We also note that the statistical error of this approach becomes negligibly small as the number of observations increases, as is also the case with our data. Where necessary, we label the exponents obtained in this way as $\rho_\theta$, $\rho_t$, and $\varepsilon_p$ to highlight that they were obtained from Eqs.~\eqref{eq:theta_l}--\eqref{eq:KramersTheory}, respectively. 

\section{Results}\label{sec:Results}

%
\subsection{Scaling exponents of random RNA sequences of varying lengths}\label{ssec:arb-len}
\begin{figure*}[!htp]
  \centering
  \includegraphics[width=\linewidth]{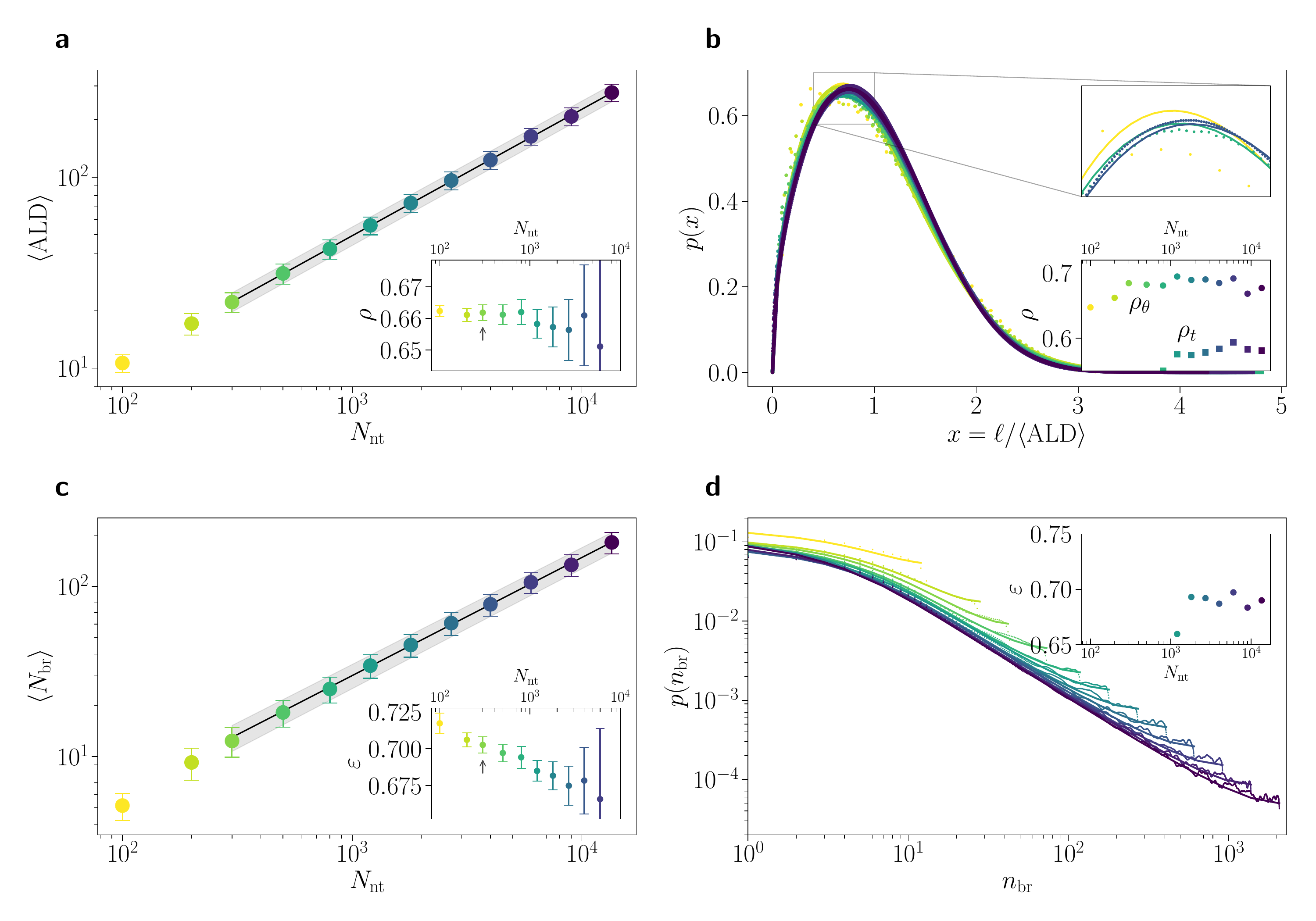}
  \caption{
  Scaling exponent $\rho$ for uniformly random RNA sequences, as obtained {\bf (a)} from the scaling of $\ALD$ with sequence length $N_\mathrm{nt}$ [Eq.~\eqref{eq:L-definition}] and {\bf (b)}~from the distributions of scaled path lengths $p(x)$, where $x=\ell/\ALD$ [Eq.~\eqref{eq:pl}]. Scaling exponent $\varepsilon$ for uniformly random RNA sequences, as obtained {\bf (c)} from the scaling of $\langle N_\mathrm{br}\rangle$ with sequence length $N_\mathrm{nt}$ [Eq.~\eqref{eq:Nbr-definition}] and {\bf (d)} from the distributions of branch weights $p(\nbr)$ [Eq.~\eqref{eq:KramersTheory}]. Insets in panels a and c show the scaling exponents obtained by fits to the data with different starting points, with the gray arrows denoting the fits over the shaded regions in the two panels. Insets in panels b and c show the scaling exponents obtained by fits to distributions at fixed RNA lengths. Colour gradients in panels b and d and their insets correspond to different sequence lengths from panels a and c. Each point in panels a and c and their corresponding line in panels b and d correspond to the average over $200$ random sequences with $500$ secondary structure thermal ensemble folds per each random sequence.
  }
\label{fig:1}
\end{figure*}

First, we take a look at how the scaling exponents $\rho$ and $\varepsilon$ can be obtained from the scaling of $\ALD$ and $\Nbr$ with RNA sequence length $N_\mathrm{nt}$, respectively.
The scaling over a range of sequence lengths for uniformly random RNAs ($f(b)=0.25$ for all $b\in\{\A,\C,\G,\U\}$) is shown in Fig.~\ref{fig:1}, panels a and c. While we consider a wide range of RNA sequence lengths ($\sim 10^2$\nobreakdash--$10^4$\,nt) to obtain the two scaling exponents, they are nonetheless {\em asymptotic} properties of the polymer (RNA tree) size. As such, their fitted values should become more reliable as the sequence length increases. This can be observed in the insets of the two panels, which show how the values of the exponents change as we change the length range of the sequences contributing to the fit, limiting the range to ever longer sequences. The exponent $\rho_{\ALD}$ quickly assumes values of around $0.66$, while the exponent $\varepsilon_{\langle N_\mathrm{br}\rangle}$ more gradually approaches similar values of around $0.68$ as shorter sequences are removed from the fit. In Sec.~\ref{ssec:rna-pol}, we further compare these values with each other as well as with the values of scaling exponents obtained from individual distributions (Sec.~\ref{ssec:fixed-len}).

As the insets of panels a and c of Fig.~\ref{fig:1} show, the range of the fit can clearly influence the value of the scaling exponent; in particular, fitting the exponents to small tree sizes ($N_\mathrm{nt}\sim10^2$) is not necessarily warranted. However, the values of the scaling exponents tend, for the most part, to remain similar within the error of the fit regardless of the fit range, and any differences start to become negligible as shorter sequences are taken out of consideration.  Sequences of length $\gtrsim 1000$~nt should thus already suffice to determine the scaling exponents.

\subsection{Robustness of the scaling exponents}\label{ssec:rob}

Now that we have determined the scaling exponents for the scaling behaviour of $\ALD$ and $\Nbr$, we wish to examine how robust these exponents are to changes in various quantities that might influence them. Specifically, we focus on the role of {\em (1)} nucleotide composition of random RNA sequences, {\em (2)} multiloop energy parameters of the secondary structure prediction software, and {\em (3)} node degree distribution of RNA trees.

\subsubsection{Nucleotide composition}
Nucleotide content can vary significantly between different biological RNAs, and has been shown to both play an important role in some of their functions as well as influence the resulting RNA structures. While our analysis primarily focuses on uniformly random RNAs to make the interpretation of our results easier, we need to verify that our conclusions remain valid also for random RNAs of different nucleotide compositions. To this purpose, we have also obtained the scaling exponents $\rho$ and $\varepsilon$ from scaling relationships for $\ALD$ and $\Nbr$ shown in panels a and c of Fig.~\ref{fig:1} for $16$ different nucleotide compositions (see Sec.~\ref{sec:Methods} and the supplementary material). The results, shown in Figs.~\ref{fig:S4} and~\ref{fig:S5} in the supplementary material, demonstrate that any observed differences in the scaling relationship are only due to changes in the prefactor of the scaling law. The scaling exponents $\rho$ and $\varepsilon$, on the other hand, remain essentially the same within the error of the fit, no matter what the nucleotide composition of random RNA is. This also means that the observations from our analysis of uniformly random RNAs can be generalized to random RNA sequences with different nucleotide compositions.

\subsubsection{Multiloop energy parameters}
Energy-based RNA secondary structure prediction depends on the accuracy of the energy parameters that are provided as an input to the folding algorithm. While the two versions of energy parameters provided by \citet{Turner2010}, Turner1999 and Turner2004, form the basis for the most commonly used secondary structure prediction software such as ViennaRNA~\cite{Lorenz2011} and RNAstructure~\cite{Mathews2006}, several improvements have also been suggested~\cite{Andronescu2010,Langdon2018}. These different sets of energy parameters noticeably differ in the energy of multiloop formation, something which is particularly relevant for large, highly branching RNA structures.

We have previously shown~\cite{Vaupotic2022} that replacing the multiloop energy parameters of the current versions of ViennaRNA (v2.0 and higher) with those of its older versions results in a significantly different node degree distributions of the resulting RNA structures. In particular, predictions made with the older version of multiloop energy parameters result in RNA structures with a much lower amount of nodes with a high degree of branching ($>4$) compared to the newer version. This is due to the energy parameter of multiloop branch formation, disfavoured in the older version of energy parameters and promoted in the newer version (see also Table~\ref{tab:multiloop} in the supplementary material).

However, when we compare the scaling of the $\ALD$ and $\Nbr$ with the RNA sequence length for structures predicted using two different sets of multiloop energy parameters (Sec.~\ref{sec:Methods}), we observe that the scaling exponents $\rho$ and $\varepsilon$ remain the same (Fig.~\ref{fig:s2} in the supplementary material). Interestingly, this implies that the scaling of RNA as a branched polymer is not strongly dependent on the particularities of the node degree distribution of its structure. This is in line with previous observations, which have found that different RNA folding algorithms and parameter sets can have a strong influence on the details of the predicted structures, but that the structure statistics were much less sensitive to them~\cite{Tacker1996}.

\subsubsection{Randomly shuffled tree topologies}
Planar tree graphs can be represented by a Pr\"ufer sequence, which has a one-to-one correspondence to a particular tree topology. By (randomly) permuting this sequence, one obtains trees with a node degree distribution identical to the original one, but with in general a different branching pattern. This approach has previously been used by Singaram {\em et al.}~\cite{Singaram2016} to study the size of ideal randomly branching polymers, where they have observed that ``RNA-like'' trees exhibit different scaling than random Pr\" ufer sequences. In Figs.~\ref{fig:prufer} and~\ref{fig:prufer2} in the supplementary material we show that Pr\"ufer-shuffling RNA trees originating from uniformly random RNA sequences indeed leads to a decrease in both scaling exponents $\rho$ and $\varepsilon$.  The exponents of Pr\"ufer-shuffled RNA trees, $\rho\approx0.53$ and $\varepsilon\approx0.59$, respectively, are indistinguishable from those of random Pr\"ufer sequences corresponding to random trees of comparable size. This shows that random trees with the same node degree distribution as random RNAs but different branching patterns scale in a significantly different fashion. Consequently, the pattern of node degree distribution alone does not suffice to explain the observed scaling relationships of RNA secondary structures. What is more, the Pr\"ufer-shuffled RNA trees appear to have even smaller size as measured by their $\MLD$ than what would be expected for a random Pr\"ufer sequence or even a random unlabelled tree with the same number of nodes.

\subsection{Scaling exponents of random RNA sequences of fixed length}\label{ssec:fixed-len}
Since the length of biological RNAs---unlike that of random RNA sequences---in general cannot be varied arbitrarily, we next demonstrate how the two scaling exponents $\rho$ and $\varepsilon$ can be determined from the distributions of path lengths $p(\ell)$ [Eq.~\eqref{eq:pl}] and branch weights $p(\nbr)$ [Eq.~\eqref{eq:KramersTheory}] in uniformly random RNA sequences of {\em fixed length}. These two distributions are shown in panels b and d of Fig.~\ref{fig:1}, respectively. To reduce the amount of noise in the data, each line represents the average over $200$ different random RNA sequences of the same length. In terms of biological RNAs, this could correspond to an average over a related group of RNAs of similar length, either by their function or by evolutionary relatedness (species, genus, \ldots).  We remark that this approach does not necessarily require all of the RNAs to have exactly the same length $N_\mathrm{nt}$, as during the mapping of RNA structure to a tree, there is already some variation in the resulting number of nodes $N$ stemming from the thermal ensemble of structures.

In general, the fits of the distributions improve as the length of the RNA increases. This is very clearly observed in the case of the distribution of branch weights, where the scaling exponent $\varepsilon$ acquires ``physical'' values ($\varepsilon \geqslant 1/2$) only for RNA lengths above $\gtrsim 800$~nt (Fig.~\ref{fig:1}d). In the case of the RdC distributions of path lengths, we can on the other hand observe that the scaling exponent $\rho$ attains a different value depending on whether it is obtained from $\theta$ or $t$ [Eqs.~\eqref{eq:theta_l} and~\eqref{eq:tl}]---both parameters of the RdC distribution (Fig.~\ref{fig:1}b; see also Fig.~\ref{fig:s3} in the supplementary material). Under certain assumptions, to which we shall return in Sec.~\ref{sec:Discussion}, one would expect that the two coefficients would be connected through the relation $\theta = 1/(t-1)$ and thus yield the same prediction for $\rho$. However, this is often not true even in the case of other branching polymers, and as Fig.~\ref{fig:s3}c also shows, certainly does not hold for random RNA sequences. This is consequently reflected in the two different predictions of $\rho$ as obtained from either $\theta$ or $t$ (Fig.~\ref{fig:1}b).

\subsection{RNA as a randomly branching polymer}\label{ssec:rna-pol}
In the previous sections, we have shown how to obtain the scaling exponents $\rho$ and $\varepsilon$ of the secondary structure of random RNA sequences in two different ways. When the sequences span a large range of lengths, the scaling exponents can be obtained from the scaling of their $\ALD$ (or $\MLD$) and $\Nbr$ with sequence length $N_\mathrm{nt}$ or, equivalently, number of nodes $N$. This leads to the scaling exponents $\rho_{\ALD}$ (or $\rho_{\MLD}$) and $\varepsilon_{\Nbr}$, respectively (Sec.~\ref{ssec:arb-len}). When the RNA sequence length cannot be varied arbitrarily, we have also shown how to obtain the same exponents through the distributions of path lengths $p(\ell)$ and branch weights $p(\nbr)$ (Sec.~\ref{ssec:fixed-len}). Since the distribution of path lengths is in general two-parametric [Eq.~\eqref{eq:pl}], this leads to two separate estimates for $\rho$, namely $\rho_\theta$ and $\rho_t$, while the distribution of branch weights gives us the exponent $\varepsilon_p$. Now we are left to examine what these scaling exponents obtained for random RNA sequences imply for the structural properties of RNA in the wider context of randomly branching polymers.

\begin{figure}[!t]
  \centering
  \includegraphics[width=\linewidth]{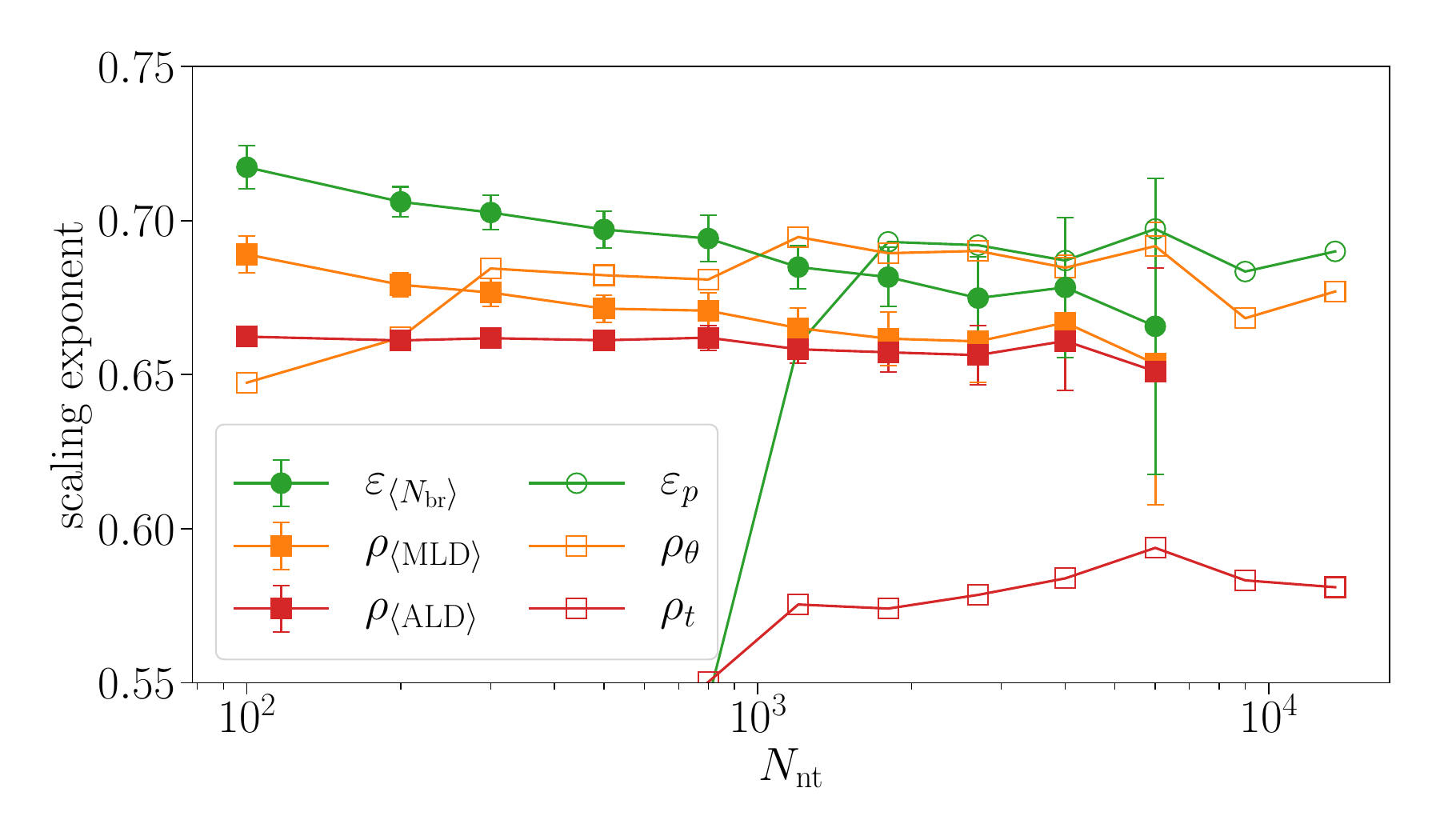}
  \caption{
  \rev{Comparison of the scaling exponents $\varepsilon$ and $\rho$ for uniformly random RNA sequences. The scaling exponents shown by full symbols were obtained from the scaling properties [Eqs.~\eqref{eq:L-definition} and~\eqref{eq:Nbr-definition}] and correspond to panels a and c of Fig.~\ref{fig:1}. The scaling exponents shown by empty symbols were obtained through fits to distributions at fixed RNA length [Eqs.~\eqref{eq:pl} and~\eqref{eq:KramersTheory}] and correspond to panels b and d of Fig.~\ref{fig:1}.}
  }
\label{fig:2}
\end{figure}

As described in Sec.~\ref{ssec:avg-pol}, one would expect for randomly branching polymers that the two topological scaling exponents $\rho$ and $\varepsilon$ will be identical, $\rho=\varepsilon$. The scaling exponents for uniformly random RNA sequences, obtained either from scaling relationships or distributions at fixed sequence length (Fig.~\ref{fig:1}) are shown together in Fig.~\ref{fig:2}. We immediately see that the exponents $\varepsilon$ and $\rho$ obtained from the scaling relationships indeed seem to converge to the same value, at least within the error bars of the fit. (Note that the sequence length in this case refers to the {\em starting sequence length} of the fit range.) When we take a look at the scaling exponents obtained from the distributions of RNA sequences of fixed length, we see that the exponents $\varepsilon$ and $\rho_\theta$ also converge to approximately the same value. This value is, furthermore, very close to the one obtained from the scaling relationships, and we can claim that we have in general
\begin{equation}
 \rho_\mathrm{RNA}\simeq\varepsilon_\mathrm{RNA}\approx0.67 .   
\end{equation}
(We remark again that for the fits to distributions, the statistical error is negligible, while the systematic error is difficult to estimate.)

A notable discrepancy, however, occurs with $\rho_t$, that is, the exponent $\rho$ obtained from the fit parameter $t$ of the RdC distribution.  The value of this scaling exponent is significantly smaller than all the other scaling exponents, even though it already appears to have converged. While it is not clear why this particular exponent does not match the other ones, we can speculate that the reason behind it lies either in the assumptions made during the derivation of the RdC curve [Eq.~\eqref{eq:q_RdC_l}] or in the Pincus blob argument used to derive the relationship between $t$ and $\rho$ [Eq.~\eqref{eq:tl}]~\cite{RosaEveraersPRE2017}. Apart from the exponent $\rho_t$, however, the scaling exponents $\rho_{\ALD}$, $\varepsilon_{\Nbr}$, $\rho_\theta$, and $\varepsilon_p$ appear to roughly obey the relationship $\rho=\varepsilon$. In this respect, RNA indeed behaves as a randomly branching polymer.

\begin{figure}[!t]
  \centering
  \includegraphics[width=.95\linewidth]{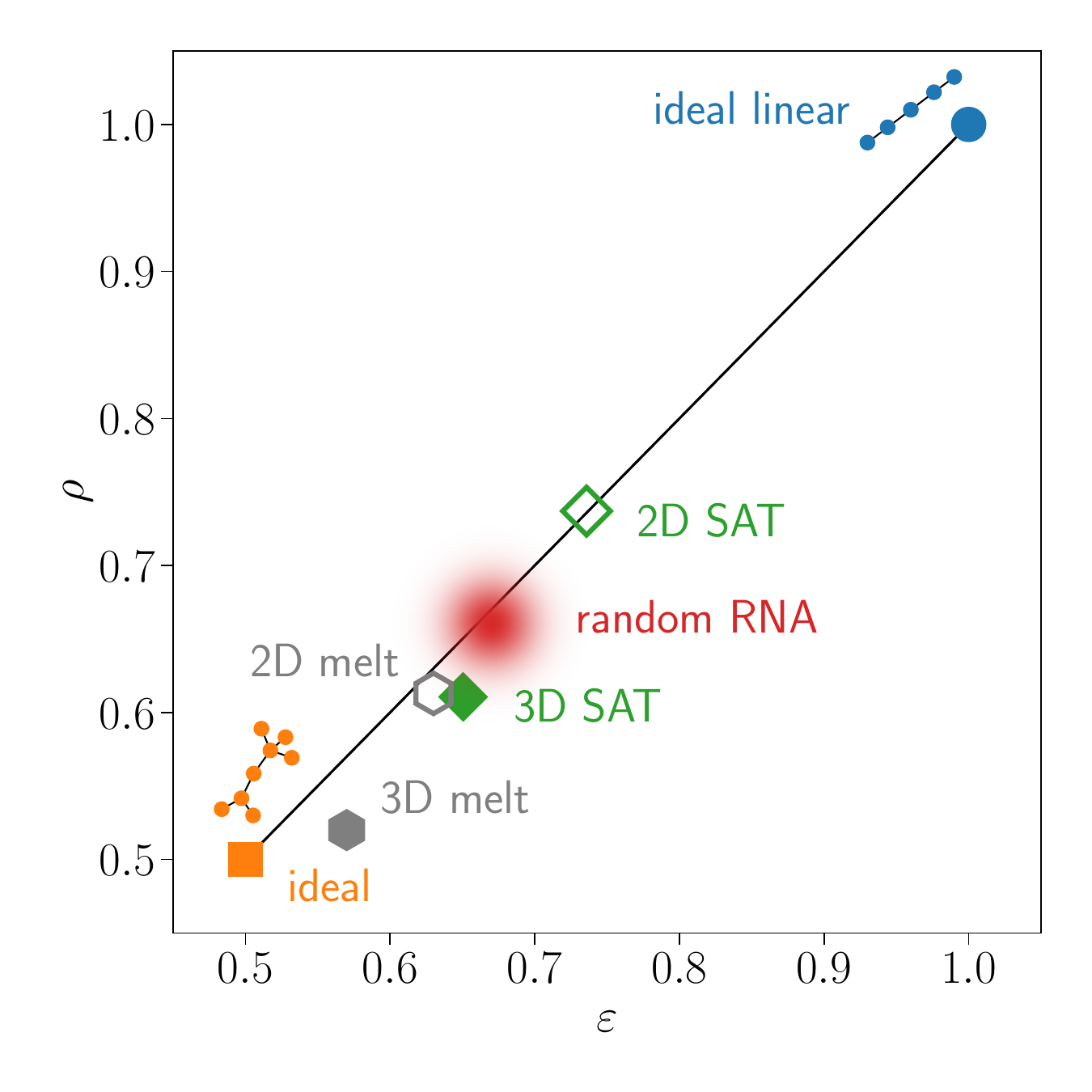}
  \caption{
  Scaling exponents $\rho$ and $\varepsilon$ for different types of branched polymers. Shown are the exact results for ideal branching and linear polymers, as well as the known computer simulation results for SATs~\cite{vanRensburg1992} and melts of randomly branching polymers~\cite{RosaEveraersJCP2016} in both 2D and 3D. Positioned in this diagram is also our result on random RNA sequences, where the smeared region depicts the estimated value of $\rho\simeq\varepsilon\approx0.67$ and the corresponding uncertainty (cf.\ Fig.~\ref{fig:2}).
  }
\label{fig:3}
\end{figure}

We can use the obtained scaling exponents to compare random RNA sequences to other types of branched polymers. Fig.~\ref{fig:3} frames our result, $\rho\simeq\varepsilon\approx0.67$, in the context of ideal linear and branched polymers, as well as self-avoiding trees (SATs)~\cite{vanRensburg1992} and melts of branching polymers~\cite{RosaEveraersJCP2016} in 2D and 3D. While we have trivially $\rho=\varepsilon=1$ for ideal linear polymers and $\rho=\varepsilon=1/2$ for ideal branched polymers, the scaling exponents of random RNAs take on values which are very similar to those of SATs in 3D. This is a particularly interesting observation since the RNA secondary structures we study are obtained using an energy-based prediction software which does not take into account any steric effects or tertiary interactions.

%

\section{Discussion}\label{sec:Discussion}
As already briefly mentioned in Sec.~\ref{sec:Theory}, the study of the connectivity of branched polymers provides an ``incomplete'' view of RNA conformations. The main reason is that---just as in the case of the more ordinary linear polymers~\cite{RubinsteinColbyBook}---how  branched conformations fold {\em in space} is a matter of a non-trivial combination between branching and the specific interactions between different monomers.

In polymer physics~\cite{RubinsteinColbyBook}, the physical characterization of the spatial conformations of any polymer ensemble can be obtained in terms of the average linear size $\langle R(N) \rangle$ defined through the root-mean-square gyration radius $\Rg$:
\begin{equation}
\label{eq:Rg-definition}
\langle R(N) \rangle \equiv \sqrt{ \langle \Rg^2\rangle } = \sqrt{ \left \langle \frac1N \sum_{i=1}^N (\vec r_i - \vec r_{\rm cm})^2 \right \rangle} \sim N^\nu ,
\end{equation}
where $\vec r_i$ is the spatial coordinate of the  $i$-th monomer and $\vec r_{\rm cm} \equiv N^{-1} \sum_{i=1}^N \vec r_i$ is the centre-of-mass of the chain. The scaling exponent $\nu$ introduced in Eq.~\eqref{eq:Rg-definition} specifies the embedding of spatial conformations {\em in space} and depends on several factors, particularly on monomer-monomer interactions and solvent conditions~\cite{Giacometti2013,Vaupotic2022}.

Contrary to what we have done for the scaling exponents $\rho$ and $\varepsilon$, the determination of the scaling exponent $\nu$ for RNA molecules is much less straightforward. \rev{The main reason, evident from Eq.~\eqref{eq:Rg-definition}, is that one needs to know either the spatial positions of all atoms or at the very least the 3D conformation of a proper coarse-grained model. Even in the latter case, the computational cost of simulating hundreds of folds for thousands of different random RNA sequences of different length remains prohibitively expensive. The possible number of RNA conformations reconstructed from experiments (for instance, NMR experiments) and computational simulations is thus still limited.} Fortunately, the theory of randomly branching polymers in the form of the classical Flory theory~\cite{FloryChemBook} comes to our aid. We now show that, under rather general assumptions, the knowledge of $\rho$ {\em implies} $\nu$~\cite{Vaupotic2022}, meaning that the branched architecture of RNA molecules and their average folding properties in space determine each other.

The Flory free energy of a randomly branching polymer or a tree with $N$ monomers is a function of the polymer mean size $\langle R\rangle$ and the average ladder distance $\langle {\rm ALD}\rangle$~\cite{GrosbergSM2014,Everaers2017}:
\begin{eqnarray}
\label{eq:FloryFreeEn}
\frac{\F}{k_{\rm B}T}
& \equiv & \frac{\F(N; \langle R\rangle, \langle {\rm ALD} \rangle)}{k_{\rm B}T} \nonumber\\
& \simeq & \frac{\langle R\rangle^2}{\langle {\rm ALD}\rangle^2} + \frac{\langle {\rm ALD}\rangle^2}{b^2N} + V(N; \langle R\rangle),
\end{eqnarray}
where $k_{\rm B}$ is the Boltzmann constant, $T$ is the temperature, $b$ is the mean bond length and $V(N; \langle R\rangle)$ is the general interaction potential between monomers which, importantly, depends only on $N$ and $\langle R\rangle$ but {\em not} on $\langle {\rm ALD}\rangle$. The sign ``$\simeq$'' in Eq.~\eqref{eq:FloryFreeEn} reminds us that the different terms in the expression are typically valid up to some numerical prefactor of $o(1)$~\cite{FloryChemBook}. When we minimize the Flory energy with respect to $\langle {\rm ALD}\rangle$~\cite{FloryChemBook,Giacometti2013,Everaers2017}, the exact form of $V$ does not matter and we easily obtain the expression connecting $\nu$ and $\rho$~\cite{Everaers2017}:
\begin{equation}\label{eq:Nu-vs-Rho-Flory}
\nu_{\rm Flory} = \frac{3\rho-1}2.
\end{equation}
As demonstrated in Ref.~\onlinecite{Everaers2017}, Eq.~\eqref{eq:Nu-vs-Rho-Flory} is in excellent agreement with all available numerical data for $\nu$ and $\rho$ in different ensembles of randomly branching polymers, both in 2D and 3D.

A strong signature for annealed connectivity of the branching topology of a molecule is the equivalence between the exponents $\rho$ and $\varepsilon$ [Eq.~\eqref{eq:Epsilon=Rho}], for which we have demonstrated that it holds true for RNA molecules as well (Figs.~\ref{fig:2} and~\ref{fig:3}). Taking all these considerations together, we propose that Eq.~\eqref{eq:Nu-vs-Rho-Flory} is applicable to RNA as well, and therefore we can use our estimated value of $\rho \simeq 2/3$ (Fig.~\ref{fig:2}) to obtain
\begin{equation}
\label{eq:NuRNA}
\nu_{\rm RNA} \simeq 1/2.
\end{equation}
This is, interestingly, the same value of the scaling exponent for randomly branching self-avoiding polymers~\cite{ParisiSourlas1981} in 3D.

We also observe that Eq.~\eqref{eq:NuRNA} contradicts the proposal by Fang {\em et al.}~\cite{Fang2011}, which was based on the Kramers' formula~\cite{Kramers1946,DaoudJoanny1981} for the radius of gyration of branching trees:
\begin{eqnarray}\label{eq:Kramers}
\langle \Rg^2 \rangle
& = &
\frac{\langle b^2\rangle}N \frac{\sum_{\nbr=0}^{N-1} \nbr \, (N-\nbr) \, \Z_{\nbr} \, \mathcal \Z_{N-\nbr}}{\sum_{\nbr=0}^{N-1} \, \Z_{\nbr} \, \Z_{N-\nbr}} \nonumber\\
& \simeq & \langle b^2\rangle N^{\rho},
\end{eqnarray}
where the last expression can be derived by using Eqs.~\eqref{eq:KramersTheory}--\eqref{eq:ansatz} with $\beta = 3/2-\varepsilon=3/2-\rho$ (see supplementary material). Here, $\langle b^2\rangle$ is the mean-square bond spatial distance of the branched RNA structure. By neglecting excluded volume interactions, $\langle b^2\rangle$ does not depend on $N$~\cite{NoteOnNuRho} and one would get $\nu=\nu_{\rm RNA}=\rho/2 \simeq 1/3$ which corresponds to the prediction by Fang {\em et al.}~\cite{Fang2011}. However, this hypothesis appears inconsistent with the measured $\rho\simeq 2/3$ as branching polymers with no volume interactions have $\rho=\varepsilon=1/2$~\cite{RubinsteinColbyBook,Everaers2017}. Since the Kramers' formula holds for both RNA molecules [Eqs.~\eqref{eq:KramersTheory}--\eqref{eq:ansatz}] and generic branching polymers~\cite{Everaers2017} alike, Eq.~\eqref{eq:Kramers} can be reconciled with it by noticing~\cite{NoteOnNuRho} that $\langle b^2\rangle \sim N^{2\nu-\rho}$. However, since this makes Eq.~\eqref{eq:Kramers} identically true, it means that the same equation cannot be used to derive $\nu$ from $\rho$.

Finally, we note that our results for $\rho$ and $\varepsilon$ are not necessarily in contradiction with the works where $\langle \Rg^2\rangle$ as a function of sequence length was determined from 3D structures of RNA molecules deposited in PDB and which led to the estimate $\nu\approx 1/3$~\cite{Thirumalai2006,Guo2022}. As described in Ref.~\onlinecite{Everaers2017}, branching polymers prepared under specific solvent conditions (for instance, in a good solvent) can be moved---by keeping the connectivity quenched!---to different solvent conditions (for instance, to a bad one) which changes the scaling exponent $\nu$ but not, of course, $\rho$~\cite{Gutin1993,Everaers2017}.
A similar process may occur during the determination of the structure of RNA molecules as well, and could be a reason for the observed discrepancy in the value of the scaling exponent $\nu$. Other potential reasons are also that the connectivity of RNAs in the PDB dataset might be quenched, or that the size distributions of these molecules is skewed towards short RNAs to the extent that one cannot observe the proper scaling exponents, which should be attainable only in the asymptotic limit of very long RNAs.

\section{Conclusions}
In this manuscript, we used the scaling theory of branching polymers (Sec.~\ref{sec:Theory}) to characterize the branching properties of ensembles of secondary structure folds of random RNAs of varying lengths and uniform composition. By ignoring pseudoknots we were able to map RNA folds to tree structures (Sec.~\ref{sec:Methods}) and obtain the scaling exponents $\rho$ and $\varepsilon$ from the length dependence of $\ALD$ and $\Nbr$, respectively (Sec.~\ref{ssec:arb-len}). Importantly, we also demonstrated how the two exponents can be determined from the distributions of path lengths and branch weights of RNA sequences with fixed length (Sec.~\ref{ssec:fixed-len}), which is particularly relevant for biological RNAs.  We found that $\rho_\mathrm{RNA}\simeq\varepsilon_\mathrm{RNA}\approx0.67$, indicating that the ensembles of RNA secondary structure folds are characterized by annealed random branching and behave as self-avoiding trees in 3D (Sec.~\ref{ssec:rna-pol}). Furthermore, we have shown that this result is robust irrespective of nucleotide composition, node degree distribution, as well as multiloop energy parameters (Sec.~\ref{ssec:rob}). Our characterization of the branched topology of RNA folds, complemented by very general polymer arguments, also allowed us to determine the scaling exponent $\nu$ for the mean spatial size of RNA molecules, where we obtained $\nu_\mathrm{RNA}=1/2$ (Sec.~\ref{sec:Discussion}). Simultaneously, we observed that other assessments of this exponent (as done by, e.g., Fang {\em et al.}~\cite{Fang2011}) implicitly assume some additional non-trivial hypotheses which do not appear to be justified based on our analysis.

Our work firmly places the branching (secondary) structure of RNA in the wider context of randomly branching polymers. In the future, the methodology developed here should be applied to biological RNAs, with viral ssRNA genomes providing a particularly good example, as the scaling exponents can be compared at the level of the viral species, genus, and family. The robustness of the scaling exponents should also be further checked by incorporating experimental reactivity data (such as SHAPE and DMS) into the prediction of RNA secondary structures. \rev{Furthermore, although previous works by \citet{ParisiSourlas1981} and \citet{LubenskyIsaacson1979} indicate that lattice animals---structures containing loops---are in the same universality class as trees, it will be interesting to study in the future how accounting for the presence of pseudoknots in RNA structures influences our results. Doing so will require a non-trivial expansion of the theory presented in this work in order to be able to study RNA structures mapped onto graphs.} We believe that a better understanding of the principles which give rise to RNA branching properties from its sequence will open up the possibility for the design of RNA sequences with desired topological properties~\cite{Jain2018,Jain2020,Rolband2022}, where the ability to design compact RNA sequences is particularly relevant in the design of vaccines~\cite{Herrero2022}.

\section*{Supplementary Material}
See the supplementary material for more details on: justification of the ansatz in Eq.~\eqref{eq:ansatz} in the main text; correlation between the quantities $N_\mathrm{nt}$, $N$, and $\widetilde{N}$; exploration of the connections between nucleotide composition, multiloop energy parameters, and Pr\"ufer shuffling with the scaling exponents $\rho$ and $\varepsilon$; and determination of the scaling exponents $\rho_t$ and $\rho_\theta$ from the distributions of path lengths.

\begin{acknowledgments}
We thank Luka Leskovec for helpful discussions. A.B.\ acknowledges support by Slovenian Research Agency (ARRS) under contract no.\ P1-0055. L.T.\ acknowledges financial support from ICSC---Centro Nazionale di Ricerca in High Performance Computing, Big Data and Quantum Computing, funded by European Union---NextGenerationEU. A.R., L.T., and A.B.\ acknowledge networking support by the COST Action CA17139 (EUTOPIA).
\end{acknowledgments}

\section*{Author declarations}

\subsection*{Conflict of interest}

The authors have no conflicts to disclose.

\subsection*{Author contributions}

Domen Vaupoti\v{c}: formal analysis; investigation; methodology; visualization; writing -- review \& editing. Angelo Rosa: conceptualization; investigation; methodology; writing -- review \& editing. Luca Tubiana: conceptualization; investigation; methodology; writing -- review \& editing. An\v{z}e Bo\v{z}i\v{c}: conceptualization; investigation; methodology; writing -- original draft; writing -- review \& editing.

\section*{Data Availability Statement}

The data that supports the findings of this study are available within the article and its supplementary material.

\bibliography{references}

\begin{thebibliography}{68}%
\makeatletter
\providecommand \@ifxundefined [1]{%
 \@ifx{#1\undefined}
}%
\providecommand \@ifnum [1]{%
 \ifnum #1\expandafter \@firstoftwo
 \else \expandafter \@secondoftwo
 \fi
}%
\providecommand \@ifx [1]{%
 \ifx #1\expandafter \@firstoftwo
 \else \expandafter \@secondoftwo
 \fi
}%
\providecommand \natexlab [1]{#1}%
\providecommand \enquote  [1]{``#1''}%
\providecommand \bibnamefont  [1]{#1}%
\providecommand \bibfnamefont [1]{#1}%
\providecommand \citenamefont [1]{#1}%
\providecommand \href@noop [0]{\@secondoftwo}%
\providecommand \href [0]{\begingroup \@sanitize@url \@href}%
\providecommand \@href[1]{\@@startlink{#1}\@@href}%
\providecommand \@@href[1]{\endgroup#1\@@endlink}%
\providecommand \@sanitize@url [0]{\catcode `\\12\catcode `\$12\catcode
  `\&12\catcode `\#12\catcode `\^12\catcode `\_12\catcode `\%12\relax}%
\providecommand \@@startlink[1]{}%
\providecommand \@@endlink[0]{}%
\providecommand \url  [0]{\begingroup\@sanitize@url \@url }%
\providecommand \@url [1]{\endgroup\@href {#1}{\urlprefix }}%
\providecommand \urlprefix  [0]{URL }%
\providecommand \Eprint [0]{\href }%
\providecommand \doibase [0]{http://dx.doi.org/}%
\providecommand \selectlanguage [0]{\@gobble}%
\providecommand \bibinfo  [0]{\@secondoftwo}%
\providecommand \bibfield  [0]{\@secondoftwo}%
\providecommand \translation [1]{[#1]}%
\providecommand \BibitemOpen [0]{}%
\providecommand \bibitemStop [0]{}%
\providecommand \bibitemNoStop [0]{.\EOS\space}%
\providecommand \EOS [0]{\spacefactor3000\relax}%
\providecommand \BibitemShut  [1]{\csname bibitem#1\endcsname}%
\let\auto@bib@innerbib\@empty
\bibitem [{\citenamefont {Kulkarni}\ and\ \citenamefont
  {Beaucage}(2006)}]{Kulkarni2006}%
  \BibitemOpen
  \bibfield  {author} {\bibinfo {author} {\bibfnamefont {A.}~\bibnamefont
  {Kulkarni}}\ and\ \bibinfo {author} {\bibfnamefont {G.}~\bibnamefont
  {Beaucage}},\ }\bibfield  {title} {\enquote {\bibinfo {title} {Quantification
  of branching in disordered materials},}\ }\href@noop {} {\bibfield  {journal}
  {\bibinfo  {journal} {J. Polym. Sci. B: Polym. Phys.}\ }\textbf {\bibinfo
  {volume} {44}},\ \bibinfo {pages} {1395--1405} (\bibinfo {year}
  {2006})}\BibitemShut {NoStop}%
\bibitem [{\citenamefont {Voit}\ and\ \citenamefont
  {Lederer}(2009)}]{Voit2009}%
  \BibitemOpen
  \bibfield  {author} {\bibinfo {author} {\bibfnamefont {B.~I.}\ \bibnamefont
  {Voit}}\ and\ \bibinfo {author} {\bibfnamefont {A.}~\bibnamefont {Lederer}},\
  }\bibfield  {title} {\enquote {\bibinfo {title} {Hyperbranched and highly
  branched polymer architectures---synthetic strategies and major
  characterization aspects},}\ }\href@noop {} {\bibfield  {journal} {\bibinfo
  {journal} {Chem. Rev.}\ }\textbf {\bibinfo {volume} {109}},\ \bibinfo {pages}
  {5924--5973} (\bibinfo {year} {2009})}\BibitemShut {NoStop}%
\bibitem [{\citenamefont {Van~der Maarel}(2007)}]{BiopolymerPhysics}%
  \BibitemOpen
  \bibfield  {author} {\bibinfo {author} {\bibfnamefont {J.~R.}\ \bibnamefont
  {Van~der Maarel}},\ }\href@noop {} {\emph {\bibinfo {title} {Introduction to
  biopolymer physics}}}\ (\bibinfo  {publisher} {World Scientific Publishing
  Company},\ \bibinfo {year} {2007})\BibitemShut {NoStop}%
\bibitem [{\citenamefont {Cook}\ and\ \citenamefont
  {Perrier}(2020)}]{Cook2020}%
  \BibitemOpen
  \bibfield  {author} {\bibinfo {author} {\bibfnamefont {A.~B.}\ \bibnamefont
  {Cook}}\ and\ \bibinfo {author} {\bibfnamefont {S.}~\bibnamefont {Perrier}},\
  }\bibfield  {title} {\enquote {\bibinfo {title} {Branched and dendritic
  polymer architectures: functional nanomaterials for therapeutic delivery},}\
  }\href@noop {} {\bibfield  {journal} {\bibinfo  {journal} {Adv. Funct.
  Mater.}\ }\textbf {\bibinfo {volume} {30}},\ \bibinfo {pages} {1901001}
  (\bibinfo {year} {2020})}\BibitemShut {NoStop}%
\bibitem [{\citenamefont {Wiedemann}\ \emph {et~al.}(2022)\citenamefont
  {Wiedemann}, \citenamefont {Kaczor}, \citenamefont {Milostan}, \citenamefont
  {Zok}, \citenamefont {Blazewicz}, \citenamefont {Szachniuk},\ and\
  \citenamefont {Antczak}}]{Wiedemann2022}%
  \BibitemOpen
  \bibfield  {author} {\bibinfo {author} {\bibfnamefont {J.}~\bibnamefont
  {Wiedemann}}, \bibinfo {author} {\bibfnamefont {J.}~\bibnamefont {Kaczor}},
  \bibinfo {author} {\bibfnamefont {M.}~\bibnamefont {Milostan}}, \bibinfo
  {author} {\bibfnamefont {T.}~\bibnamefont {Zok}}, \bibinfo {author}
  {\bibfnamefont {J.}~\bibnamefont {Blazewicz}}, \bibinfo {author}
  {\bibfnamefont {M.}~\bibnamefont {Szachniuk}}, \ and\ \bibinfo {author}
  {\bibfnamefont {M.}~\bibnamefont {Antczak}},\ }\bibfield  {title} {\enquote
  {\bibinfo {title} {Rnaloops: a database of {RNA} multiloops},}\ }\href@noop
  {} {\bibfield  {journal} {\bibinfo  {journal} {Bioinformatics}\ } (\bibinfo
  {year} {2022})}\BibitemShut {NoStop}%
\bibitem [{\citenamefont {Boerneke}, \citenamefont {Ehrhardt},\ and\
  \citenamefont {Weeks}(2019)}]{Boerneke2019}%
  \BibitemOpen
  \bibfield  {author} {\bibinfo {author} {\bibfnamefont {M.~A.}\ \bibnamefont
  {Boerneke}}, \bibinfo {author} {\bibfnamefont {J.~E.}\ \bibnamefont
  {Ehrhardt}}, \ and\ \bibinfo {author} {\bibfnamefont {K.~M.}\ \bibnamefont
  {Weeks}},\ }\bibfield  {title} {\enquote {\bibinfo {title} {Physical and
  functional analysis of viral {RNA} genomes by {SHAPE}},}\ }\href@noop {}
  {\bibfield  {journal} {\bibinfo  {journal} {Annu. Rev. Virol.}\ }\textbf
  {\bibinfo {volume} {6}},\ \bibinfo {pages} {93} (\bibinfo {year}
  {2019})}\BibitemShut {NoStop}%
\bibitem [{\citenamefont {Wan}\ \emph {et~al.}(2022)\citenamefont {Wan},
  \citenamefont {Adams}, \citenamefont {Lindenbach},\ and\ \citenamefont
  {Pyle}}]{Wan2022}%
  \BibitemOpen
  \bibfield  {author} {\bibinfo {author} {\bibfnamefont {H.}~\bibnamefont
  {Wan}}, \bibinfo {author} {\bibfnamefont {R.~L.}\ \bibnamefont {Adams}},
  \bibinfo {author} {\bibfnamefont {B.~D.}\ \bibnamefont {Lindenbach}}, \ and\
  \bibinfo {author} {\bibfnamefont {A.~M.}\ \bibnamefont {Pyle}},\ }\bibfield
  {title} {\enquote {\bibinfo {title} {The in vivo and in vitro architecture of
  the {Hepatitis C} virus {RNA} genome uncovers functional {RNA} secondary and
  tertiary structures},}\ }\href@noop {} {\bibfield  {journal} {\bibinfo
  {journal} {J. Virol.}\ }\textbf {\bibinfo {volume} {96}},\ \bibinfo {pages}
  {e01946--21} (\bibinfo {year} {2022})}\BibitemShut {NoStop}%
\bibitem [{\citenamefont {Yoffe}\ \emph {et~al.}(2008)\citenamefont {Yoffe},
  \citenamefont {Prinsen}, \citenamefont {Gopal}, \citenamefont {Knobler},
  \citenamefont {Gelbart},\ and\ \citenamefont {Ben-Shaul}}]{Yoffe2008}%
  \BibitemOpen
  \bibfield  {author} {\bibinfo {author} {\bibfnamefont {A.~M.}\ \bibnamefont
  {Yoffe}}, \bibinfo {author} {\bibfnamefont {P.}~\bibnamefont {Prinsen}},
  \bibinfo {author} {\bibfnamefont {A.}~\bibnamefont {Gopal}}, \bibinfo
  {author} {\bibfnamefont {C.~M.}\ \bibnamefont {Knobler}}, \bibinfo {author}
  {\bibfnamefont {W.~M.}\ \bibnamefont {Gelbart}}, \ and\ \bibinfo {author}
  {\bibfnamefont {A.}~\bibnamefont {Ben-Shaul}},\ }\bibfield  {title} {\enquote
  {\bibinfo {title} {Predicting the sizes of large {RNA} molecules},}\
  }\href@noop {} {\bibfield  {journal} {\bibinfo  {journal} {Proc. Natl. Acad.
  Sci. USA}\ }\textbf {\bibinfo {volume} {105}},\ \bibinfo {pages}
  {16153--16158} (\bibinfo {year} {2008})}\BibitemShut {NoStop}%
\bibitem [{\citenamefont {Tubiana}\ \emph {et~al.}(2015)\citenamefont
  {Tubiana}, \citenamefont {Bo{\v{z}}i{\v{c}}}, \citenamefont {Micheletti},\
  and\ \citenamefont {Podgornik}}]{Tubiana2015}%
  \BibitemOpen
  \bibfield  {author} {\bibinfo {author} {\bibfnamefont {L.}~\bibnamefont
  {Tubiana}}, \bibinfo {author} {\bibfnamefont {A.}~\bibnamefont
  {Bo{\v{z}}i{\v{c}}}}, \bibinfo {author} {\bibfnamefont {C.}~\bibnamefont
  {Micheletti}}, \ and\ \bibinfo {author} {\bibfnamefont {R.}~\bibnamefont
  {Podgornik}},\ }\bibfield  {title} {\enquote {\bibinfo {title} {Synonymous
  mutations reduce genome compactness in icosahedral {ssRNA} viruses},}\
  }\href@noop {} {\bibfield  {journal} {\bibinfo  {journal} {Biophys. J.}\
  }\textbf {\bibinfo {volume} {108}},\ \bibinfo {pages} {194--202} (\bibinfo
  {year} {2015})}\BibitemShut {NoStop}%
\bibitem [{\citenamefont {Singaram}\ \emph {et~al.}(2015)\citenamefont
  {Singaram}, \citenamefont {Garmann}, \citenamefont {Knobler}, \citenamefont
  {Gelbart},\ and\ \citenamefont {Ben-Shaul}}]{Singaram2015}%
  \BibitemOpen
  \bibfield  {author} {\bibinfo {author} {\bibfnamefont {S.~W.}\ \bibnamefont
  {Singaram}}, \bibinfo {author} {\bibfnamefont {R.~F.}\ \bibnamefont
  {Garmann}}, \bibinfo {author} {\bibfnamefont {C.~M.}\ \bibnamefont
  {Knobler}}, \bibinfo {author} {\bibfnamefont {W.~M.}\ \bibnamefont
  {Gelbart}}, \ and\ \bibinfo {author} {\bibfnamefont {A.}~\bibnamefont
  {Ben-Shaul}},\ }\bibfield  {title} {\enquote {\bibinfo {title} {Role of {RNA}
  branchedness in the competition for viral capsid proteins},}\ }\href@noop {}
  {\bibfield  {journal} {\bibinfo  {journal} {J. Phys. Chem. B}\ }\textbf
  {\bibinfo {volume} {119}},\ \bibinfo {pages} {13991--14002} (\bibinfo {year}
  {2015})}\BibitemShut {NoStop}%
\bibitem [{\citenamefont {Garmann}\ \emph {et~al.}(2016)\citenamefont
  {Garmann}, \citenamefont {Comas-Garcia}, \citenamefont {Knobler},\ and\
  \citenamefont {Gelbart}}]{Garmann2016}%
  \BibitemOpen
  \bibfield  {author} {\bibinfo {author} {\bibfnamefont {R.~F.}\ \bibnamefont
  {Garmann}}, \bibinfo {author} {\bibfnamefont {M.}~\bibnamefont
  {Comas-Garcia}}, \bibinfo {author} {\bibfnamefont {C.~M.}\ \bibnamefont
  {Knobler}}, \ and\ \bibinfo {author} {\bibfnamefont {W.~M.}\ \bibnamefont
  {Gelbart}},\ }\bibfield  {title} {\enquote {\bibinfo {title} {Physical
  principles in the self-assembly of a simple spherical virus},}\ }\href@noop
  {} {\bibfield  {journal} {\bibinfo  {journal} {Acc. Chem. Res.}\ }\textbf
  {\bibinfo {volume} {49}},\ \bibinfo {pages} {48--55} (\bibinfo {year}
  {2016})}\BibitemShut {NoStop}%
\bibitem [{\citenamefont {Beren}\ \emph {et~al.}(2017)\citenamefont {Beren},
  \citenamefont {Dreesens}, \citenamefont {Liu}, \citenamefont {Knobler},\ and\
  \citenamefont {Gelbart}}]{Beren2017}%
  \BibitemOpen
  \bibfield  {author} {\bibinfo {author} {\bibfnamefont {C.}~\bibnamefont
  {Beren}}, \bibinfo {author} {\bibfnamefont {L.~L.}\ \bibnamefont {Dreesens}},
  \bibinfo {author} {\bibfnamefont {K.~N.}\ \bibnamefont {Liu}}, \bibinfo
  {author} {\bibfnamefont {C.~M.}\ \bibnamefont {Knobler}}, \ and\ \bibinfo
  {author} {\bibfnamefont {W.~M.}\ \bibnamefont {Gelbart}},\ }\bibfield
  {title} {\enquote {\bibinfo {title} {The effect of {RNA} secondary structure
  on the self-assembly of viral capsids},}\ }\href@noop {} {\bibfield
  {journal} {\bibinfo  {journal} {Biophys. J.}\ }\textbf {\bibinfo {volume}
  {113}},\ \bibinfo {pages} {339--347} (\bibinfo {year} {2017})}\BibitemShut
  {NoStop}%
\bibitem [{\citenamefont {{Bo\v{z}i\v{c}}}\ \emph {et~al.}(2018)\citenamefont
  {{Bo\v{z}i\v{c}}}, \citenamefont {Micheletti}, \citenamefont {Podgornik},\
  and\ \citenamefont {Tubiana}}]{Bozic2018}%
  \BibitemOpen
  \bibfield  {author} {\bibinfo {author} {\bibfnamefont {A.}~\bibnamefont
  {{Bo\v{z}i\v{c}}}}, \bibinfo {author} {\bibfnamefont {C.}~\bibnamefont
  {Micheletti}}, \bibinfo {author} {\bibfnamefont {R.}~\bibnamefont
  {Podgornik}}, \ and\ \bibinfo {author} {\bibfnamefont {L.}~\bibnamefont
  {Tubiana}},\ }\bibfield  {title} {\enquote {\bibinfo {title} {Compactness of
  viral genomes: effect of disperse and localized random mutations},}\
  }\href@noop {} {\bibfield  {journal} {\bibinfo  {journal} {J. Phys. Condens.
  Matter}\ }\textbf {\bibinfo {volume} {30}},\ \bibinfo {pages} {084006}
  (\bibinfo {year} {2018})}\BibitemShut {NoStop}%
\bibitem [{\citenamefont {Marichal}\ \emph {et~al.}(2021)\citenamefont
  {Marichal}, \citenamefont {Gargowitsch}, \citenamefont {Rubim}, \citenamefont
  {Sizun}, \citenamefont {Kra}, \citenamefont {Bressanelli}, \citenamefont
  {Dong}, \citenamefont {Panahandeh}, \citenamefont {Zandi},\ and\
  \citenamefont {Tresset}}]{Marichal2021}%
  \BibitemOpen
  \bibfield  {author} {\bibinfo {author} {\bibfnamefont {L.}~\bibnamefont
  {Marichal}}, \bibinfo {author} {\bibfnamefont {L.}~\bibnamefont
  {Gargowitsch}}, \bibinfo {author} {\bibfnamefont {R.~L.}\ \bibnamefont
  {Rubim}}, \bibinfo {author} {\bibfnamefont {C.}~\bibnamefont {Sizun}},
  \bibinfo {author} {\bibfnamefont {K.}~\bibnamefont {Kra}}, \bibinfo {author}
  {\bibfnamefont {S.}~\bibnamefont {Bressanelli}}, \bibinfo {author}
  {\bibfnamefont {Y.}~\bibnamefont {Dong}}, \bibinfo {author} {\bibfnamefont
  {S.}~\bibnamefont {Panahandeh}}, \bibinfo {author} {\bibfnamefont
  {R.}~\bibnamefont {Zandi}}, \ and\ \bibinfo {author} {\bibfnamefont
  {G.}~\bibnamefont {Tresset}},\ }\bibfield  {title} {\enquote {\bibinfo
  {title} {Relationships between {RNA} topology and nucleocapsid structure in a
  model icosahedral virus},}\ }\href@noop {} {\bibfield  {journal} {\bibinfo
  {journal} {Biophys. J.}\ }\textbf {\bibinfo {volume} {120}},\ \bibinfo
  {pages} {3925--3936} (\bibinfo {year} {2021})}\BibitemShut {NoStop}%
\bibitem [{\citenamefont {Zandi}\ and\ \citenamefont {Van~der
  Schoot}(2009)}]{Zandi2009}%
  \BibitemOpen
  \bibfield  {author} {\bibinfo {author} {\bibfnamefont {R.}~\bibnamefont
  {Zandi}}\ and\ \bibinfo {author} {\bibfnamefont {P.}~\bibnamefont {Van~der
  Schoot}},\ }\bibfield  {title} {\enquote {\bibinfo {title} {Size regulation
  of {ss-RNA} viruses},}\ }\href@noop {} {\bibfield  {journal} {\bibinfo
  {journal} {Biophys. J.}\ }\textbf {\bibinfo {volume} {96}},\ \bibinfo {pages}
  {9--20} (\bibinfo {year} {2009})}\BibitemShut {NoStop}%
\bibitem [{\citenamefont {van~der Schoot}\ and\ \citenamefont
  {Zandi}(2013)}]{vds2013}%
  \BibitemOpen
  \bibfield  {author} {\bibinfo {author} {\bibfnamefont {P.}~\bibnamefont
  {van~der Schoot}}\ and\ \bibinfo {author} {\bibfnamefont {R.}~\bibnamefont
  {Zandi}},\ }\bibfield  {title} {\enquote {\bibinfo {title} {Impact of the
  topology of viral {RNAs} on their encapsulation by virus coat proteins},}\
  }\href@noop {} {\bibfield  {journal} {\bibinfo  {journal} {J. Biol. Phys.}\
  }\textbf {\bibinfo {volume} {39}},\ \bibinfo {pages} {289--299} (\bibinfo
  {year} {2013})}\BibitemShut {NoStop}%
\bibitem [{\citenamefont {Wagner}, \citenamefont {Erdemci-Tandogan},\ and\
  \citenamefont {Zandi}(2015)}]{Zandi2015}%
  \BibitemOpen
  \bibfield  {author} {\bibinfo {author} {\bibfnamefont {J.}~\bibnamefont
  {Wagner}}, \bibinfo {author} {\bibfnamefont {G.}~\bibnamefont
  {Erdemci-Tandogan}}, \ and\ \bibinfo {author} {\bibfnamefont
  {R.}~\bibnamefont {Zandi}},\ }\bibfield  {title} {\enquote {\bibinfo {title}
  {Adsorption of annealed branched polymers on curved surfaces},}\ }\href@noop
  {} {\bibfield  {journal} {\bibinfo  {journal} {J. Phys. Condens. Matter}\
  }\textbf {\bibinfo {volume} {27}},\ \bibinfo {pages} {495101} (\bibinfo
  {year} {2015})}\BibitemShut {NoStop}%
\bibitem [{\citenamefont {Erdemci-Tandogan}\ \emph {et~al.}(2014)\citenamefont
  {Erdemci-Tandogan}, \citenamefont {Wagner}, \citenamefont {Van Der~Schoot},
  \citenamefont {Podgornik},\ and\ \citenamefont {Zandi}}]{Gonca2014}%
  \BibitemOpen
  \bibfield  {author} {\bibinfo {author} {\bibfnamefont {G.}~\bibnamefont
  {Erdemci-Tandogan}}, \bibinfo {author} {\bibfnamefont {J.}~\bibnamefont
  {Wagner}}, \bibinfo {author} {\bibfnamefont {P.}~\bibnamefont {Van
  Der~Schoot}}, \bibinfo {author} {\bibfnamefont {R.}~\bibnamefont
  {Podgornik}}, \ and\ \bibinfo {author} {\bibfnamefont {R.}~\bibnamefont
  {Zandi}},\ }\bibfield  {title} {\enquote {\bibinfo {title} {{RNA} topology
  remolds electrostatic stabilization of viruses},}\ }\href@noop {} {\bibfield
  {journal} {\bibinfo  {journal} {Phys. Rev. E}\ }\textbf {\bibinfo {volume}
  {89}},\ \bibinfo {pages} {032707} (\bibinfo {year} {2014})}\BibitemShut
  {NoStop}%
\bibitem [{\citenamefont {Erdemci-Tandogan}\ \emph {et~al.}(2016)\citenamefont
  {Erdemci-Tandogan}, \citenamefont {Wagner}, \citenamefont {van~der Schoot},
  \citenamefont {Podgornik},\ and\ \citenamefont {Zandi}}]{Erdemci2016}%
  \BibitemOpen
  \bibfield  {author} {\bibinfo {author} {\bibfnamefont {G.}~\bibnamefont
  {Erdemci-Tandogan}}, \bibinfo {author} {\bibfnamefont {J.}~\bibnamefont
  {Wagner}}, \bibinfo {author} {\bibfnamefont {P.}~\bibnamefont {van~der
  Schoot}}, \bibinfo {author} {\bibfnamefont {R.}~\bibnamefont {Podgornik}}, \
  and\ \bibinfo {author} {\bibfnamefont {R.}~\bibnamefont {Zandi}},\ }\bibfield
   {title} {\enquote {\bibinfo {title} {Effects of {RNA} branching on the
  electrostatic stabilization of viruses},}\ }\href@noop {} {\bibfield
  {journal} {\bibinfo  {journal} {Phys. Rev. E}\ }\textbf {\bibinfo {volume}
  {94}},\ \bibinfo {pages} {022408} (\bibinfo {year} {2016})}\BibitemShut
  {NoStop}%
\bibitem [{\citenamefont {Gopal}\ \emph {et~al.}(2014)\citenamefont {Gopal},
  \citenamefont {Egecioglu}, \citenamefont {Yoffe}, \citenamefont {Ben-Shaul},
  \citenamefont {Rao}, \citenamefont {Knobler},\ and\ \citenamefont
  {Gelbart}}]{Gopal2014}%
  \BibitemOpen
  \bibfield  {author} {\bibinfo {author} {\bibfnamefont {A.}~\bibnamefont
  {Gopal}}, \bibinfo {author} {\bibfnamefont {D.~E.}\ \bibnamefont
  {Egecioglu}}, \bibinfo {author} {\bibfnamefont {A.~M.}\ \bibnamefont
  {Yoffe}}, \bibinfo {author} {\bibfnamefont {A.}~\bibnamefont {Ben-Shaul}},
  \bibinfo {author} {\bibfnamefont {A.~L.}\ \bibnamefont {Rao}}, \bibinfo
  {author} {\bibfnamefont {C.~M.}\ \bibnamefont {Knobler}}, \ and\ \bibinfo
  {author} {\bibfnamefont {W.~M.}\ \bibnamefont {Gelbart}},\ }\bibfield
  {title} {\enquote {\bibinfo {title} {Viral {RNAs} are unusually compact},}\
  }\href@noop {} {\bibfield  {journal} {\bibinfo  {journal} {PLoS One}\
  }\textbf {\bibinfo {volume} {9}},\ \bibinfo {pages} {e105875} (\bibinfo
  {year} {2014})}\BibitemShut {NoStop}%
\bibitem [{\citenamefont {Garmann}\ \emph {et~al.}(2015)\citenamefont
  {Garmann}, \citenamefont {Gopal}, \citenamefont {Athavale}, \citenamefont
  {Knobler}, \citenamefont {Gelbart},\ and\ \citenamefont
  {Harvey}}]{Garmann2015}%
  \BibitemOpen
  \bibfield  {author} {\bibinfo {author} {\bibfnamefont {R.~F.}\ \bibnamefont
  {Garmann}}, \bibinfo {author} {\bibfnamefont {A.}~\bibnamefont {Gopal}},
  \bibinfo {author} {\bibfnamefont {S.~S.}\ \bibnamefont {Athavale}}, \bibinfo
  {author} {\bibfnamefont {C.~M.}\ \bibnamefont {Knobler}}, \bibinfo {author}
  {\bibfnamefont {W.~M.}\ \bibnamefont {Gelbart}}, \ and\ \bibinfo {author}
  {\bibfnamefont {S.~C.}\ \bibnamefont {Harvey}},\ }\bibfield  {title}
  {\enquote {\bibinfo {title} {Visualizing the global secondary structure of a
  viral {RNA} genome with cryo-electron microscopy},}\ }\href@noop {}
  {\bibfield  {journal} {\bibinfo  {journal} {RNA}\ }\textbf {\bibinfo {volume}
  {21}},\ \bibinfo {pages} {877--886} (\bibinfo {year} {2015})}\BibitemShut
  {NoStop}%
\bibitem [{\citenamefont {Gan}, \citenamefont {Pasquali},\ and\ \citenamefont
  {Schlick}(2003)}]{Gan2003}%
  \BibitemOpen
  \bibfield  {author} {\bibinfo {author} {\bibfnamefont {H.~H.}\ \bibnamefont
  {Gan}}, \bibinfo {author} {\bibfnamefont {S.}~\bibnamefont {Pasquali}}, \
  and\ \bibinfo {author} {\bibfnamefont {T.}~\bibnamefont {Schlick}},\
  }\bibfield  {title} {\enquote {\bibinfo {title} {Exploring the repertoire of
  {RNA} secondary motifs using graph theory; implications for {RNA} design},}\
  }\href@noop {} {\bibfield  {journal} {\bibinfo  {journal} {Nucleic Acids
  Res.}\ }\textbf {\bibinfo {volume} {31}},\ \bibinfo {pages} {2926--2943}
  (\bibinfo {year} {2003})}\BibitemShut {NoStop}%
\bibitem [{\citenamefont {Schlick}(2018)}]{Schlick2018}%
  \BibitemOpen
  \bibfield  {author} {\bibinfo {author} {\bibfnamefont {T.}~\bibnamefont
  {Schlick}},\ }\bibfield  {title} {\enquote {\bibinfo {title} {Adventures with
  {RNA} graphs},}\ }\href@noop {} {\bibfield  {journal} {\bibinfo  {journal}
  {Methods}\ }\textbf {\bibinfo {volume} {143}},\ \bibinfo {pages} {16--33}
  (\bibinfo {year} {2018})}\BibitemShut {NoStop}%
\bibitem [{\citenamefont {Vaupoti\v{c}}\ \emph {et~al.}(2022)\citenamefont
  {Vaupoti\v{c}}, \citenamefont {Rosa}, \citenamefont {Podgornik},
  \citenamefont {Tubiana},\ and\ \citenamefont
  {{Bo\v{z}i\v{c}}}}]{Vaupotic2022}%
  \BibitemOpen
  \bibfield  {author} {\bibinfo {author} {\bibfnamefont {D.}~\bibnamefont
  {Vaupoti\v{c}}}, \bibinfo {author} {\bibfnamefont {A.}~\bibnamefont {Rosa}},
  \bibinfo {author} {\bibfnamefont {R.}~\bibnamefont {Podgornik}}, \bibinfo
  {author} {\bibfnamefont {L.}~\bibnamefont {Tubiana}}, \ and\ \bibinfo
  {author} {\bibfnamefont {A.}~\bibnamefont {{Bo\v{z}i\v{c}}}},\ }\href@noop {}
  {\enquote {\bibinfo {title} {Viral {RNA} as a branched polymer},}\ }
  (\bibinfo {year} {2022}),\ \Eprint {http://arxiv.org/abs/arXiv:2212.00829}
  {arXiv:2212.00829 [physics.bio-ph]} \BibitemShut {NoStop}%
\bibitem [{\citenamefont {Borodavka}\ \emph {et~al.}(2016)\citenamefont
  {Borodavka}, \citenamefont {Singaram}, \citenamefont {Stockley},
  \citenamefont {Gelbart}, \citenamefont {Ben-Shaul},\ and\ \citenamefont
  {Tuma}}]{Borodavka2016}%
  \BibitemOpen
  \bibfield  {author} {\bibinfo {author} {\bibfnamefont {A.}~\bibnamefont
  {Borodavka}}, \bibinfo {author} {\bibfnamefont {S.~W.}\ \bibnamefont
  {Singaram}}, \bibinfo {author} {\bibfnamefont {P.~G.}\ \bibnamefont
  {Stockley}}, \bibinfo {author} {\bibfnamefont {W.~M.}\ \bibnamefont
  {Gelbart}}, \bibinfo {author} {\bibfnamefont {A.}~\bibnamefont {Ben-Shaul}},
  \ and\ \bibinfo {author} {\bibfnamefont {R.}~\bibnamefont {Tuma}},\
  }\bibfield  {title} {\enquote {\bibinfo {title} {Sizes of long {RNA}
  molecules are determined by the branching patterns of their secondary
  structures},}\ }\href@noop {} {\bibfield  {journal} {\bibinfo  {journal}
  {Biophys. J.}\ }\textbf {\bibinfo {volume} {111}},\ \bibinfo {pages}
  {2077--2085} (\bibinfo {year} {2016})}\BibitemShut {NoStop}%
\bibitem [{\citenamefont {Flory}(1953)}]{FloryChemBook}%
  \BibitemOpen
  \bibfield  {author} {\bibinfo {author} {\bibfnamefont {P.~J.}\ \bibnamefont
  {Flory}},\ }\href@noop {} {\emph {\bibinfo {title} {Principles of Polymer
  Chemistry}}}\ (\bibinfo  {publisher} {Cornell University Press},\ \bibinfo
  {address} {Ithaca (NY)},\ \bibinfo {year} {1953})\BibitemShut {NoStop}%
\bibitem [{\citenamefont {Bhattacharjee}, \citenamefont {Giacometti},\ and\
  \citenamefont {Maritan}(2013)}]{Giacometti2013}%
  \BibitemOpen
  \bibfield  {author} {\bibinfo {author} {\bibfnamefont {S.~M.}\ \bibnamefont
  {Bhattacharjee}}, \bibinfo {author} {\bibfnamefont {A.}~\bibnamefont
  {Giacometti}}, \ and\ \bibinfo {author} {\bibfnamefont {A.}~\bibnamefont
  {Maritan}},\ }\bibfield  {title} {\enquote {\bibinfo {title} {{Flory} theory
  for polymers},}\ }\href@noop {} {\bibfield  {journal} {\bibinfo  {journal}
  {J. Phys. Cond. Matter}\ }\textbf {\bibinfo {volume} {25}},\ \bibinfo {pages}
  {503101} (\bibinfo {year} {2013})}\BibitemShut {NoStop}%
\bibitem [{\citenamefont {Everaers}\ \emph {et~al.}(2017)\citenamefont
  {Everaers}, \citenamefont {Grosberg}, \citenamefont {Rubinstein},\ and\
  \citenamefont {Rosa}}]{Everaers2017}%
  \BibitemOpen
  \bibfield  {author} {\bibinfo {author} {\bibfnamefont {R.}~\bibnamefont
  {Everaers}}, \bibinfo {author} {\bibfnamefont {A.~Y.}\ \bibnamefont
  {Grosberg}}, \bibinfo {author} {\bibfnamefont {M.}~\bibnamefont
  {Rubinstein}}, \ and\ \bibinfo {author} {\bibfnamefont {A.}~\bibnamefont
  {Rosa}},\ }\bibfield  {title} {\enquote {\bibinfo {title} {{Flory} theory of
  randomly branched polymers},}\ }\href@noop {} {\bibfield  {journal} {\bibinfo
   {journal} {Soft Matter}\ }\textbf {\bibinfo {volume} {13}},\ \bibinfo
  {pages} {1223--1234} (\bibinfo {year} {2017})}\BibitemShut {NoStop}%
\bibitem [{\citenamefont {Rubinstein}\ and\ \citenamefont
  {Colby}(2003)}]{RubinsteinColbyBook}%
  \BibitemOpen
  \bibfield  {author} {\bibinfo {author} {\bibfnamefont {M.}~\bibnamefont
  {Rubinstein}}\ and\ \bibinfo {author} {\bibfnamefont {R.~H.}\ \bibnamefont
  {Colby}},\ }\href@noop {} {\emph {\bibinfo {title} {Polymer Physics}}}\
  (\bibinfo  {publisher} {Oxford University Press},\ \bibinfo {address} {New
  York},\ \bibinfo {year} {2003})\BibitemShut {NoStop}%
\bibitem [{\citenamefont {Wang}(2017)}]{Wang2017}%
  \BibitemOpen
  \bibfield  {author} {\bibinfo {author} {\bibfnamefont {Z.-G.}\ \bibnamefont
  {Wang}},\ }\bibfield  {title} {\enquote {\bibinfo {title} {50th anniversary
  perspective: Polymer conformation---a pedagogical review},}\ }\href@noop {}
  {\bibfield  {journal} {\bibinfo  {journal} {Macromolecules}\ }\textbf
  {\bibinfo {volume} {50}},\ \bibinfo {pages} {9073--9114} (\bibinfo {year}
  {2017})}\BibitemShut {NoStop}%
\bibitem [{\citenamefont {Fang}, \citenamefont {Gelbart},\ and\ \citenamefont
  {Ben-Shaul}(2011)}]{Fang2011}%
  \BibitemOpen
  \bibfield  {author} {\bibinfo {author} {\bibfnamefont {L.~T.}\ \bibnamefont
  {Fang}}, \bibinfo {author} {\bibfnamefont {W.~M.}\ \bibnamefont {Gelbart}}, \
  and\ \bibinfo {author} {\bibfnamefont {A.}~\bibnamefont {Ben-Shaul}},\
  }\bibfield  {title} {\enquote {\bibinfo {title} {The size of {RNA} as an
  ideal branched polymer},}\ }\href@noop {} {\bibfield  {journal} {\bibinfo
  {journal} {J. Chem. Phys.}\ }\textbf {\bibinfo {volume} {135}},\ \bibinfo
  {pages} {10B616} (\bibinfo {year} {2011})}\BibitemShut {NoStop}%
\bibitem [{\citenamefont {Kramers}(1946)}]{Kramers1946}%
  \BibitemOpen
  \bibfield  {author} {\bibinfo {author} {\bibfnamefont {H.~A.}\ \bibnamefont
  {Kramers}},\ }\bibfield  {title} {\enquote {\bibinfo {title} {The behavior of
  macromolecules in inhomogeneous flow},}\ }\href@noop {} {\bibfield  {journal}
  {\bibinfo  {journal} {J. Chem. Phys.}\ }\textbf {\bibinfo {volume} {14}},\
  \bibinfo {pages} {415--424} (\bibinfo {year} {1946})}\BibitemShut {NoStop}%
\bibitem [{\citenamefont {Lubensky}\ and\ \citenamefont
  {Isaacson}(1979)}]{LubenskyIsaacson1979}%
  \BibitemOpen
  \bibfield  {author} {\bibinfo {author} {\bibfnamefont {T.}~\bibnamefont
  {Lubensky}}\ and\ \bibinfo {author} {\bibfnamefont {J.}~\bibnamefont
  {Isaacson}},\ }\bibfield  {title} {\enquote {\bibinfo {title} {Statistics of
  lattice animals and dilute branched polymers},}\ }\href@noop {} {\bibfield
  {journal} {\bibinfo  {journal} {Phys. Rev. A}\ }\textbf {\bibinfo {volume}
  {20}},\ \bibinfo {pages} {2130--2146} (\bibinfo {year} {1979})}\BibitemShut
  {NoStop}%
\bibitem [{\citenamefont {Daoud}\ and\ \citenamefont
  {Joanny}(1981)}]{DaoudJoanny1981}%
  \BibitemOpen
  \bibfield  {author} {\bibinfo {author} {\bibfnamefont {M.}~\bibnamefont
  {Daoud}}\ and\ \bibinfo {author} {\bibfnamefont {J.~F.}\ \bibnamefont
  {Joanny}},\ }\bibfield  {title} {\enquote {\bibinfo {title} {Conformation of
  branched polymers},}\ }\href@noop {} {\bibfield  {journal} {\bibinfo
  {journal} {J. Physique}\ }\textbf {\bibinfo {volume} {42}},\ \bibinfo {pages}
  {1359--1371} (\bibinfo {year} {1981})}\BibitemShut {NoStop}%
\bibitem [{\citenamefont {Van~Rensburg}\ and\ \citenamefont
  {Madras}(1992)}]{vanRensburg1992}%
  \BibitemOpen
  \bibfield  {author} {\bibinfo {author} {\bibfnamefont {E.~J.}\ \bibnamefont
  {Van~Rensburg}}\ and\ \bibinfo {author} {\bibfnamefont {N.}~\bibnamefont
  {Madras}},\ }\bibfield  {title} {\enquote {\bibinfo {title} {A nonlocal
  {Monte Carlo} algorithm for lattice trees},}\ }\href@noop {} {\bibfield
  {journal} {\bibinfo  {journal} {J. Phys. A: Math. Theor.}\ }\textbf {\bibinfo
  {volume} {25}},\ \bibinfo {pages} {303} (\bibinfo {year} {1992})}\BibitemShut
  {NoStop}%
\bibitem [{\citenamefont {Grosberg}\ and\ \citenamefont
  {Bruinsma}(2018)}]{Grosberg2018}%
  \BibitemOpen
  \bibfield  {author} {\bibinfo {author} {\bibfnamefont {A.~Y.}\ \bibnamefont
  {Grosberg}}\ and\ \bibinfo {author} {\bibfnamefont {R.}~\bibnamefont
  {Bruinsma}},\ }\bibfield  {title} {\enquote {\bibinfo {title} {Confining
  annealed branched polymers inside spherical capsids},}\ }\href@noop {}
  {\bibfield  {journal} {\bibinfo  {journal} {J. Biol. Phys.}\ }\textbf
  {\bibinfo {volume} {44}},\ \bibinfo {pages} {133--145} (\bibinfo {year}
  {2018})}\BibitemShut {NoStop}%
\bibitem [{\citenamefont {Kelly}, \citenamefont {Grosberg},\ and\ \citenamefont
  {Bruinsma}(2016)}]{Kelly2016}%
  \BibitemOpen
  \bibfield  {author} {\bibinfo {author} {\bibfnamefont {J.}~\bibnamefont
  {Kelly}}, \bibinfo {author} {\bibfnamefont {A.~Y.}\ \bibnamefont {Grosberg}},
  \ and\ \bibinfo {author} {\bibfnamefont {R.}~\bibnamefont {Bruinsma}},\
  }\bibfield  {title} {\enquote {\bibinfo {title} {Sequence dependence of viral
  {RNA} encapsidation},}\ }\href@noop {} {\bibfield  {journal} {\bibinfo
  {journal} {J. Phys. Chem. B}\ }\textbf {\bibinfo {volume} {120}},\ \bibinfo
  {pages} {6038--6050} (\bibinfo {year} {2016})}\BibitemShut {NoStop}%
\bibitem [{\citenamefont {Higgs}(1993)}]{Higgs1993}%
  \BibitemOpen
  \bibfield  {author} {\bibinfo {author} {\bibfnamefont {P.~G.}\ \bibnamefont
  {Higgs}},\ }\bibfield  {title} {\enquote {\bibinfo {title} {{RNA} secondary
  structure: a comparison of real and random sequences},}\ }\href@noop {}
  {\bibfield  {journal} {\bibinfo  {journal} {J. Phys., I}\ }\textbf {\bibinfo
  {volume} {3}},\ \bibinfo {pages} {43--59} (\bibinfo {year}
  {1993})}\BibitemShut {NoStop}%
\bibitem [{\citenamefont {Bundschuh}\ and\ \citenamefont
  {Hwa}(2002)}]{Bundschuh2002}%
  \BibitemOpen
  \bibfield  {author} {\bibinfo {author} {\bibfnamefont {R.}~\bibnamefont
  {Bundschuh}}\ and\ \bibinfo {author} {\bibfnamefont {T.}~\bibnamefont
  {Hwa}},\ }\bibfield  {title} {\enquote {\bibinfo {title} {Statistical
  mechanics of secondary structures formed by random {RNA} sequences},}\
  }\href@noop {} {\bibfield  {journal} {\bibinfo  {journal} {Phys. Rev. E}\
  }\textbf {\bibinfo {volume} {65}},\ \bibinfo {pages} {031903} (\bibinfo
  {year} {2002})}\BibitemShut {NoStop}%
\bibitem [{\citenamefont {Schultes}\ \emph {et~al.}(2005)\citenamefont
  {Schultes}, \citenamefont {Spasic}, \citenamefont {Mohanty},\ and\
  \citenamefont {Bartel}}]{Schultes2005}%
  \BibitemOpen
  \bibfield  {author} {\bibinfo {author} {\bibfnamefont {E.~A.}\ \bibnamefont
  {Schultes}}, \bibinfo {author} {\bibfnamefont {A.}~\bibnamefont {Spasic}},
  \bibinfo {author} {\bibfnamefont {U.}~\bibnamefont {Mohanty}}, \ and\
  \bibinfo {author} {\bibfnamefont {D.~P.}\ \bibnamefont {Bartel}},\ }\bibfield
   {title} {\enquote {\bibinfo {title} {Compact and ordered collapse of
  randomly generated {RNA} sequences},}\ }\href@noop {} {\bibfield  {journal}
  {\bibinfo  {journal} {Nat. Struct. Mol. Biol.}\ }\textbf {\bibinfo {volume}
  {12}},\ \bibinfo {pages} {1130--1136} (\bibinfo {year} {2005})}\BibitemShut
  {NoStop}%
\bibitem [{\citenamefont {Clote}\ \emph {et~al.}(2005)\citenamefont {Clote},
  \citenamefont {Ferr{\'e}}, \citenamefont {Kranakis},\ and\ \citenamefont
  {Krizanc}}]{Clote2005}%
  \BibitemOpen
  \bibfield  {author} {\bibinfo {author} {\bibfnamefont {P.}~\bibnamefont
  {Clote}}, \bibinfo {author} {\bibfnamefont {F.}~\bibnamefont {Ferr{\'e}}},
  \bibinfo {author} {\bibfnamefont {E.}~\bibnamefont {Kranakis}}, \ and\
  \bibinfo {author} {\bibfnamefont {D.}~\bibnamefont {Krizanc}},\ }\bibfield
  {title} {\enquote {\bibinfo {title} {Structural {RNA} has lower folding
  energy than random {RNA} of the same dinucleotide frequency},}\ }\href@noop
  {} {\bibfield  {journal} {\bibinfo  {journal} {RNA}\ }\textbf {\bibinfo
  {volume} {11}},\ \bibinfo {pages} {578--591} (\bibinfo {year}
  {2005})}\BibitemShut {NoStop}%
\bibitem [{\citenamefont {Chizzolini}\ \emph {et~al.}(2019)\citenamefont
  {Chizzolini}, \citenamefont {Passalacqua}, \citenamefont {Oumais},
  \citenamefont {Dingilian}, \citenamefont {Szostak},\ and\ \citenamefont
  {Luptak}}]{Chizzolini2019}%
  \BibitemOpen
  \bibfield  {author} {\bibinfo {author} {\bibfnamefont {F.}~\bibnamefont
  {Chizzolini}}, \bibinfo {author} {\bibfnamefont {L.~F.}\ \bibnamefont
  {Passalacqua}}, \bibinfo {author} {\bibfnamefont {M.}~\bibnamefont {Oumais}},
  \bibinfo {author} {\bibfnamefont {A.~I.}\ \bibnamefont {Dingilian}}, \bibinfo
  {author} {\bibfnamefont {J.~W.}\ \bibnamefont {Szostak}}, \ and\ \bibinfo
  {author} {\bibfnamefont {A.}~\bibnamefont {Luptak}},\ }\bibfield  {title}
  {\enquote {\bibinfo {title} {Large phenotypic enhancement of structured
  random {RNA} pools},}\ }\href@noop {} {\bibfield  {journal} {\bibinfo
  {journal} {J. Am. Chem. Soc.}\ }\textbf {\bibinfo {volume} {142}},\ \bibinfo
  {pages} {1941--1951} (\bibinfo {year} {2019})}\BibitemShut {NoStop}%
\bibitem [{\citenamefont {Rosa}\ and\ \citenamefont
  {Everaers}(2016{\natexlab{a}})}]{RosaEveraersJPA2016}%
  \BibitemOpen
  \bibfield  {author} {\bibinfo {author} {\bibfnamefont {A.}~\bibnamefont
  {Rosa}}\ and\ \bibinfo {author} {\bibfnamefont {R.}~\bibnamefont
  {Everaers}},\ }\bibfield  {title} {\enquote {\bibinfo {title} {Computer
  simulations of randomly branching polymers: annealed versus quenched
  branching structures},}\ }\href@noop {} {\bibfield  {journal} {\bibinfo
  {journal} {J. Phys. A Math. Theor.}\ }\textbf {\bibinfo {volume} {49}},\
  \bibinfo {pages} {345001} (\bibinfo {year} {2016}{\natexlab{a}})}\BibitemShut
  {NoStop}%
\bibitem [{\citenamefont {Rosa}\ and\ \citenamefont
  {Everaers}(2016{\natexlab{b}})}]{RosaEveraersJCP2016}%
  \BibitemOpen
  \bibfield  {author} {\bibinfo {author} {\bibfnamefont {A.}~\bibnamefont
  {Rosa}}\ and\ \bibinfo {author} {\bibfnamefont {R.}~\bibnamefont
  {Everaers}},\ }\bibfield  {title} {\enquote {\bibinfo {title} {Computer
  simulations of melts of randomly branching polymers},}\ }\href@noop {}
  {\bibfield  {journal} {\bibinfo  {journal} {J. Chem. Phys.}\ }\textbf
  {\bibinfo {volume} {145}},\ \bibinfo {pages} {164906} (\bibinfo {year}
  {2016}{\natexlab{b}})}\BibitemShut {NoStop}%
\bibitem [{\citenamefont {Rosa}\ and\ \citenamefont
  {Everaers}(2017)}]{RosaEveraersPRE2017}%
  \BibitemOpen
  \bibfield  {author} {\bibinfo {author} {\bibfnamefont {A.}~\bibnamefont
  {Rosa}}\ and\ \bibinfo {author} {\bibfnamefont {R.}~\bibnamefont
  {Everaers}},\ }\bibfield  {title} {\enquote {\bibinfo {title} {Beyond {Flory}
  theory: Distribution functions for interacting lattice trees},}\ }\href@noop
  {} {\bibfield  {journal} {\bibinfo  {journal} {Phys. Rev. E}\ }\textbf
  {\bibinfo {volume} {95}},\ \bibinfo {pages} {012117} (\bibinfo {year}
  {2017})}\BibitemShut {NoStop}%
\bibitem [{The()}]{Theta-ell-WhyNotation}%
  \BibitemOpen
  \href@noop {} {}\bibinfo {note} {In the original work of
  \citet{RosaEveraersPRE2017}, the exponents $\theta$ and $t$ and the related
  numerical constants $C$ and $K$ bear the subscript $\ell$ to distinguish them
  from analogous quantities appearing in other distribution functions. Here, to
  lighten up the notation, we drop this subscript.}\BibitemShut {Stop}%
\bibitem [{\citenamefont {Woods}\ \emph {et~al.}(2017)\citenamefont {Woods},
  \citenamefont {Lackey}, \citenamefont {Williams}, \citenamefont {Dokholyan},
  \citenamefont {Gotz},\ and\ \citenamefont {Laederach}}]{Woods2017}%
  \BibitemOpen
  \bibfield  {author} {\bibinfo {author} {\bibfnamefont {C.~T.}\ \bibnamefont
  {Woods}}, \bibinfo {author} {\bibfnamefont {L.}~\bibnamefont {Lackey}},
  \bibinfo {author} {\bibfnamefont {B.}~\bibnamefont {Williams}}, \bibinfo
  {author} {\bibfnamefont {N.~V.}\ \bibnamefont {Dokholyan}}, \bibinfo {author}
  {\bibfnamefont {D.}~\bibnamefont {Gotz}}, \ and\ \bibinfo {author}
  {\bibfnamefont {A.}~\bibnamefont {Laederach}},\ }\bibfield  {title} {\enquote
  {\bibinfo {title} {Comparative visualization of the {RNA} suboptimal
  conformational ensemble in vivo},}\ }\href@noop {} {\bibfield  {journal}
  {\bibinfo  {journal} {Biophys. J.}\ }\textbf {\bibinfo {volume} {113}},\
  \bibinfo {pages} {290--301} (\bibinfo {year} {2017})}\BibitemShut {NoStop}%
\bibitem [{\citenamefont {Lorenz}\ \emph {et~al.}(2011)\citenamefont {Lorenz},
  \citenamefont {Bernhart}, \citenamefont {H{\"o}ner~zu Siederdissen},
  \citenamefont {Tafer}, \citenamefont {Flamm}, \citenamefont {Stadler},\ and\
  \citenamefont {Hofacker}}]{Lorenz2011}%
  \BibitemOpen
  \bibfield  {author} {\bibinfo {author} {\bibfnamefont {R.}~\bibnamefont
  {Lorenz}}, \bibinfo {author} {\bibfnamefont {S.~H.}\ \bibnamefont
  {Bernhart}}, \bibinfo {author} {\bibfnamefont {C.}~\bibnamefont {H{\"o}ner~zu
  Siederdissen}}, \bibinfo {author} {\bibfnamefont {H.}~\bibnamefont {Tafer}},
  \bibinfo {author} {\bibfnamefont {C.}~\bibnamefont {Flamm}}, \bibinfo
  {author} {\bibfnamefont {P.~F.}\ \bibnamefont {Stadler}}, \ and\ \bibinfo
  {author} {\bibfnamefont {I.~L.}\ \bibnamefont {Hofacker}},\ }\bibfield
  {title} {\enquote {\bibinfo {title} {{ViennaRNA} package 2.0},}\ }\href@noop
  {} {\bibfield  {journal} {\bibinfo  {journal} {Algorithms Mol. Biol.}\
  }\textbf {\bibinfo {volume} {6}},\ \bibinfo {pages} {1--14} (\bibinfo {year}
  {2011})}\BibitemShut {NoStop}%
\bibitem [{\citenamefont {Singaram}, \citenamefont {Gopal},\ and\ \citenamefont
  {Ben-Shaul}(2016)}]{Singaram2016}%
  \BibitemOpen
  \bibfield  {author} {\bibinfo {author} {\bibfnamefont {S.~W.}\ \bibnamefont
  {Singaram}}, \bibinfo {author} {\bibfnamefont {A.}~\bibnamefont {Gopal}}, \
  and\ \bibinfo {author} {\bibfnamefont {A.}~\bibnamefont {Ben-Shaul}},\
  }\bibfield  {title} {\enquote {\bibinfo {title} {A {Pr\"ufer}-sequence based
  algorithm for calculating the size of ideal randomly branched polymers},}\
  }\href@noop {} {\bibfield  {journal} {\bibinfo  {journal} {J. Phys. Chem. B}\
  }\textbf {\bibinfo {volume} {120}},\ \bibinfo {pages} {6231--6237} (\bibinfo
  {year} {2016})}\BibitemShut {NoStop}%
\bibitem [{\citenamefont {Poznanovi{\'c}}\ \emph {et~al.}(2021)\citenamefont
  {Poznanovi{\'c}}, \citenamefont {Wood}, \citenamefont {Cloer},\ and\
  \citenamefont {Heitsch}}]{Poznanovic2021}%
  \BibitemOpen
  \bibfield  {author} {\bibinfo {author} {\bibfnamefont {S.}~\bibnamefont
  {Poznanovi{\'c}}}, \bibinfo {author} {\bibfnamefont {C.}~\bibnamefont
  {Wood}}, \bibinfo {author} {\bibfnamefont {M.}~\bibnamefont {Cloer}}, \ and\
  \bibinfo {author} {\bibfnamefont {C.}~\bibnamefont {Heitsch}},\ }\bibfield
  {title} {\enquote {\bibinfo {title} {Improving {RNA} branching predictions:
  advances and limitations},}\ }\href@noop {} {\bibfield  {journal} {\bibinfo
  {journal} {Genes}\ }\textbf {\bibinfo {volume} {12}},\ \bibinfo {pages} {469}
  (\bibinfo {year} {2021})}\BibitemShut {NoStop}%
\bibitem [{\citenamefont {Ward}\ \emph {et~al.}(2017)\citenamefont {Ward},
  \citenamefont {Datta}, \citenamefont {Wise},\ and\ \citenamefont
  {Mathews}}]{Ward2017}%
  \BibitemOpen
  \bibfield  {author} {\bibinfo {author} {\bibfnamefont {M.}~\bibnamefont
  {Ward}}, \bibinfo {author} {\bibfnamefont {A.}~\bibnamefont {Datta}},
  \bibinfo {author} {\bibfnamefont {M.}~\bibnamefont {Wise}}, \ and\ \bibinfo
  {author} {\bibfnamefont {D.~H.}\ \bibnamefont {Mathews}},\ }\bibfield
  {title} {\enquote {\bibinfo {title} {Advanced multi-loop algorithms for {RNA}
  secondary structure prediction reveal that the simplest model is best},}\
  }\href@noop {} {\bibfield  {journal} {\bibinfo  {journal} {Nucleic Acids
  Res.}\ }\textbf {\bibinfo {volume} {45}},\ \bibinfo {pages} {8541--8550}
  (\bibinfo {year} {2017})}\BibitemShut {NoStop}%
\bibitem [{\citenamefont {Poznanovi{\'c}}\ \emph {et~al.}(2020)\citenamefont
  {Poznanovi{\'c}}, \citenamefont {Barrera-Cruz}, \citenamefont {Kirkpatrick},
  \citenamefont {Ielusic},\ and\ \citenamefont {Heitsch}}]{Poznanovic2020}%
  \BibitemOpen
  \bibfield  {author} {\bibinfo {author} {\bibfnamefont {S.}~\bibnamefont
  {Poznanovi{\'c}}}, \bibinfo {author} {\bibfnamefont {F.}~\bibnamefont
  {Barrera-Cruz}}, \bibinfo {author} {\bibfnamefont {A.}~\bibnamefont
  {Kirkpatrick}}, \bibinfo {author} {\bibfnamefont {M.}~\bibnamefont
  {Ielusic}}, \ and\ \bibinfo {author} {\bibfnamefont {C.}~\bibnamefont
  {Heitsch}},\ }\bibfield  {title} {\enquote {\bibinfo {title} {The challenge
  of {RNA} branching prediction: a parametric analysis of multiloop initiation
  under thermodynamic optimization},}\ }\href@noop {} {\bibfield  {journal}
  {\bibinfo  {journal} {J. Struct. Biol.}\ }\textbf {\bibinfo {volume} {210}},\
  \bibinfo {pages} {107475} (\bibinfo {year} {2020})}\BibitemShut {NoStop}%
\bibitem [{\citenamefont {Clauset}, \citenamefont {Shalizi},\ and\
  \citenamefont {Newman}(2009)}]{Clauset2009}%
  \BibitemOpen
  \bibfield  {author} {\bibinfo {author} {\bibfnamefont {A.}~\bibnamefont
  {Clauset}}, \bibinfo {author} {\bibfnamefont {C.~R.}\ \bibnamefont
  {Shalizi}}, \ and\ \bibinfo {author} {\bibfnamefont {M.~E.}\ \bibnamefont
  {Newman}},\ }\bibfield  {title} {\enquote {\bibinfo {title} {{Power-law
  distributions in empirical data}},}\ }\href@noop {} {\bibfield  {journal}
  {\bibinfo  {journal} {SIAM Rev.}\ }\textbf {\bibinfo {volume} {51}},\
  \bibinfo {pages} {661--703} (\bibinfo {year} {2009})}\BibitemShut {NoStop}%
\bibitem [{\citenamefont {Turner}\ and\ \citenamefont
  {Mathews}(2010)}]{Turner2010}%
  \BibitemOpen
  \bibfield  {author} {\bibinfo {author} {\bibfnamefont {D.~H.}\ \bibnamefont
  {Turner}}\ and\ \bibinfo {author} {\bibfnamefont {D.~H.}\ \bibnamefont
  {Mathews}},\ }\bibfield  {title} {\enquote {\bibinfo {title} {{NNDB}: the
  nearest neighbor parameter database for predicting stability of nucleic acid
  secondary structure},}\ }\href@noop {} {\bibfield  {journal} {\bibinfo
  {journal} {Nucleic Acids Res.}\ }\textbf {\bibinfo {volume} {38}},\ \bibinfo
  {pages} {D280--D282} (\bibinfo {year} {2010})}\BibitemShut {NoStop}%
\bibitem [{\citenamefont {Mathews}\ and\ \citenamefont
  {Turner}(2006)}]{Mathews2006}%
  \BibitemOpen
  \bibfield  {author} {\bibinfo {author} {\bibfnamefont {D.~H.}\ \bibnamefont
  {Mathews}}\ and\ \bibinfo {author} {\bibfnamefont {D.~H.}\ \bibnamefont
  {Turner}},\ }\bibfield  {title} {\enquote {\bibinfo {title} {Prediction of
  {RNA} secondary structure by free energy minimization},}\ }\href@noop {}
  {\bibfield  {journal} {\bibinfo  {journal} {Curr. Op. Struct. Biol.}\
  }\textbf {\bibinfo {volume} {16}},\ \bibinfo {pages} {270--278} (\bibinfo
  {year} {2006})}\BibitemShut {NoStop}%
\bibitem [{\citenamefont {Andronescu}\ \emph {et~al.}(2010)\citenamefont
  {Andronescu}, \citenamefont {Condon}, \citenamefont {Hoos}, \citenamefont
  {Mathews},\ and\ \citenamefont {Murphy}}]{Andronescu2010}%
  \BibitemOpen
  \bibfield  {author} {\bibinfo {author} {\bibfnamefont {M.}~\bibnamefont
  {Andronescu}}, \bibinfo {author} {\bibfnamefont {A.}~\bibnamefont {Condon}},
  \bibinfo {author} {\bibfnamefont {H.~H.}\ \bibnamefont {Hoos}}, \bibinfo
  {author} {\bibfnamefont {D.~H.}\ \bibnamefont {Mathews}}, \ and\ \bibinfo
  {author} {\bibfnamefont {K.~P.}\ \bibnamefont {Murphy}},\ }\bibfield  {title}
  {\enquote {\bibinfo {title} {Computational approaches for {RNA} energy
  parameter estimation},}\ }\href@noop {} {\bibfield  {journal} {\bibinfo
  {journal} {RNA}\ }\textbf {\bibinfo {volume} {16}},\ \bibinfo {pages}
  {2304--2318} (\bibinfo {year} {2010})}\BibitemShut {NoStop}%
\bibitem [{\citenamefont {Langdon}, \citenamefont {Petke},\ and\ \citenamefont
  {Lorenz}(2018)}]{Langdon2018}%
  \BibitemOpen
  \bibfield  {author} {\bibinfo {author} {\bibfnamefont {W.~B.}\ \bibnamefont
  {Langdon}}, \bibinfo {author} {\bibfnamefont {J.}~\bibnamefont {Petke}}, \
  and\ \bibinfo {author} {\bibfnamefont {R.}~\bibnamefont {Lorenz}},\
  }\bibfield  {title} {\enquote {\bibinfo {title} {Evolving better {RNAfold}
  structure prediction},}\ }in\ \href@noop {} {\emph {\bibinfo {booktitle}
  {European Conference on Genetic Programming}}}\ (\bibinfo {year} {2018})\
  pp.\ \bibinfo {pages} {220--236}\BibitemShut {NoStop}%
\bibitem [{\citenamefont {Tacker}\ \emph {et~al.}(1996)\citenamefont {Tacker},
  \citenamefont {Stadler}, \citenamefont {Bornberg-Bauer}, \citenamefont
  {Hofacker},\ and\ \citenamefont {Schuster}}]{Tacker1996}%
  \BibitemOpen
  \bibfield  {author} {\bibinfo {author} {\bibfnamefont {M.}~\bibnamefont
  {Tacker}}, \bibinfo {author} {\bibfnamefont {P.~F.}\ \bibnamefont {Stadler}},
  \bibinfo {author} {\bibfnamefont {E.~G.}\ \bibnamefont {Bornberg-Bauer}},
  \bibinfo {author} {\bibfnamefont {I.~L.}\ \bibnamefont {Hofacker}}, \ and\
  \bibinfo {author} {\bibfnamefont {P.}~\bibnamefont {Schuster}},\ }\bibfield
  {title} {\enquote {\bibinfo {title} {Algorithm independent properties of
  {RNA} secondary structure predictions},}\ }\href@noop {} {\bibfield
  {journal} {\bibinfo  {journal} {Eur. Biophys. J.}\ }\textbf {\bibinfo
  {volume} {25}},\ \bibinfo {pages} {115--130} (\bibinfo {year}
  {1996})}\BibitemShut {NoStop}%
\bibitem [{\citenamefont {Grosberg}(2014)}]{GrosbergSM2014}%
  \BibitemOpen
  \bibfield  {author} {\bibinfo {author} {\bibfnamefont {A.~Y.}\ \bibnamefont
  {Grosberg}},\ }\bibfield  {title} {\enquote {\bibinfo {title} {Annealed
  lattice animal model and {Flory} theory for the melt of non-concatenated
  rings: towards the physics of crumpling},}\ }\href@noop {} {\bibfield
  {journal} {\bibinfo  {journal} {Soft Matter}\ }\textbf {\bibinfo {volume}
  {10}},\ \bibinfo {pages} {560--565} (\bibinfo {year} {2014})}\BibitemShut
  {NoStop}%
\bibitem [{\citenamefont {Parisi}\ and\ \citenamefont
  {Sourlas}(1981)}]{ParisiSourlas1981}%
  \BibitemOpen
  \bibfield  {author} {\bibinfo {author} {\bibfnamefont {G.}~\bibnamefont
  {Parisi}}\ and\ \bibinfo {author} {\bibfnamefont {N.}~\bibnamefont
  {Sourlas}},\ }\bibfield  {title} {\enquote {\bibinfo {title} {Critical
  behavior of branched polymers and the {Lee-Yang} edge singularity},}\
  }\href@noop {} {\bibfield  {journal} {\bibinfo  {journal} {Phys. Rev. Lett.}\
  }\textbf {\bibinfo {volume} {46}},\ \bibinfo {pages} {871--874} (\bibinfo
  {year} {1981})}\BibitemShut {NoStop}%
\bibitem [{Not()}]{NoteOnNuRho}%
  \BibitemOpen
  \href@noop {} {}\bibinfo {note} {According to the Kramers'
  theorem~\cite{Kramers1946,DaoudJoanny1981,RubinsteinColbyBook}, each bond
  contributes $\simeq \langle b^2\rangle$ to the mean-square radius of
  gyration. In terms of the notation of this paper, $\langle b^2\rangle$ can be
  defined as the ratio between the mean-square end-to-end distance between tree
  paths with average ladder distance $\langle {\rm ALD}\rangle$ and the last
  one, i.e., \begin{eqnarray}\label{eq:Define<b^2>} \langle b^2\rangle & \simeq
  & \frac{\langle R^2(\langle {\rm ALD}\rangle)\rangle}{\langle {\rm
  ALD}\rangle/b} \simeq b^2 \frac{(\langle {\rm ALD}\rangle/b)^{2\nu_{\rm
  path}}}{\langle {\rm ALD}\rangle/b} \nonumber\\ & = & b^2 (\langle {\rm
  ALD}\rangle / b)^{2\nu_{\rm path}-1} \, , \end{eqnarray} where $b$ is the
  mean bond length as in Eq.~\eqref{eq:FloryFreeEn} and $\nu_{\rm path} = \nu /
  \rho$~\cite{Everaers2017} is the scaling exponent for the spatial structure
  of tree linear paths. For non-interacting trees with $\nu=1/4$ and
  $\rho=1/2$~\cite{Everaers2017}, $\nu_{\rm path}=1/2$ and $\langle b^2\rangle
  \simeq b^2$ does not depend on $N$. Conversely, for interacting trees we do
  have $\langle {\rm ALD} \rangle /b \sim N^{\rho}$
  [Eq.~\eqref{eq:L-definition}] and Eq.~\eqref{eq:Define<b^2>} then implies
  that $\langle b^2\rangle \simeq b^2 N^{2\nu-\rho}$.}\BibitemShut {Stop}%
\bibitem [{\citenamefont {Hyeon}, \citenamefont {Dima},\ and\ \citenamefont
  {Thirumalai}(2006)}]{Thirumalai2006}%
  \BibitemOpen
  \bibfield  {author} {\bibinfo {author} {\bibfnamefont {C.}~\bibnamefont
  {Hyeon}}, \bibinfo {author} {\bibfnamefont {R.~I.}\ \bibnamefont {Dima}}, \
  and\ \bibinfo {author} {\bibfnamefont {D.}~\bibnamefont {Thirumalai}},\
  }\bibfield  {title} {\enquote {\bibinfo {title} {Size, shape, and flexibility
  of {RNA} structures},}\ }\href@noop {} {\bibfield  {journal} {\bibinfo
  {journal} {J. Chem. Phys.}\ }\textbf {\bibinfo {volume} {125}},\ \bibinfo
  {pages} {194905} (\bibinfo {year} {2006})}\BibitemShut {NoStop}%
\bibitem [{\citenamefont {Guo}\ \emph {et~al.}(2022)\citenamefont {Guo},
  \citenamefont {Yuan}, \citenamefont {Tan}, \citenamefont {Zhang},\ and\
  \citenamefont {Shi}}]{Guo2022}%
  \BibitemOpen
  \bibfield  {author} {\bibinfo {author} {\bibfnamefont {Z.-H.}\ \bibnamefont
  {Guo}}, \bibinfo {author} {\bibfnamefont {L.}~\bibnamefont {Yuan}}, \bibinfo
  {author} {\bibfnamefont {Y.-L.}\ \bibnamefont {Tan}}, \bibinfo {author}
  {\bibfnamefont {B.-G.}\ \bibnamefont {Zhang}}, \ and\ \bibinfo {author}
  {\bibfnamefont {Y.-Z.}\ \bibnamefont {Shi}},\ }\bibfield  {title} {\enquote
  {\bibinfo {title} {{RNAStat}: An integrated tool for statistical analysis of
  {RNA} {3D} structures},}\ }\href@noop {} {\bibfield  {journal} {\bibinfo
  {journal} {Front. Bioinform.}\ }\textbf {\bibinfo {volume} {1}},\ \bibinfo
  {pages} {809082} (\bibinfo {year} {2022})}\BibitemShut {NoStop}%
\bibitem [{\citenamefont {Gutin}, \citenamefont {Grosberg},\ and\ \citenamefont
  {Shakhnovich}(1993)}]{Gutin1993}%
  \BibitemOpen
  \bibfield  {author} {\bibinfo {author} {\bibfnamefont {A.~M.}\ \bibnamefont
  {Gutin}}, \bibinfo {author} {\bibfnamefont {A.~Y.}\ \bibnamefont {Grosberg}},
  \ and\ \bibinfo {author} {\bibfnamefont {E.~I.}\ \bibnamefont
  {Shakhnovich}},\ }\bibfield  {title} {\enquote {\bibinfo {title} {Polymers
  with annealed and quenched branchings belong to different universality
  classes},}\ }\href@noop {} {\bibfield  {journal} {\bibinfo  {journal}
  {Macromolecules}\ }\textbf {\bibinfo {volume} {26}},\ \bibinfo {pages}
  {1293--1295} (\bibinfo {year} {1993})}\BibitemShut {NoStop}%
\bibitem [{\citenamefont {Jain}\ \emph {et~al.}(2018)\citenamefont {Jain},
  \citenamefont {Laederach}, \citenamefont {Ramos},\ and\ \citenamefont
  {Schlick}}]{Jain2018}%
  \BibitemOpen
  \bibfield  {author} {\bibinfo {author} {\bibfnamefont {S.}~\bibnamefont
  {Jain}}, \bibinfo {author} {\bibfnamefont {A.}~\bibnamefont {Laederach}},
  \bibinfo {author} {\bibfnamefont {S.~B.}\ \bibnamefont {Ramos}}, \ and\
  \bibinfo {author} {\bibfnamefont {T.}~\bibnamefont {Schlick}},\ }\bibfield
  {title} {\enquote {\bibinfo {title} {A pipeline for computational design of
  novel {RNA}-like topologies},}\ }\href@noop {} {\bibfield  {journal}
  {\bibinfo  {journal} {Nucleic Acids Res.}\ }\textbf {\bibinfo {volume}
  {46}},\ \bibinfo {pages} {7040--7051} (\bibinfo {year} {2018})}\BibitemShut
  {NoStop}%
\bibitem [{\citenamefont {Jain}, \citenamefont {Tao},\ and\ \citenamefont
  {Schlick}(2020)}]{Jain2020}%
  \BibitemOpen
  \bibfield  {author} {\bibinfo {author} {\bibfnamefont {S.}~\bibnamefont
  {Jain}}, \bibinfo {author} {\bibfnamefont {Y.}~\bibnamefont {Tao}}, \ and\
  \bibinfo {author} {\bibfnamefont {T.}~\bibnamefont {Schlick}},\ }\bibfield
  {title} {\enquote {\bibinfo {title} {Inverse folding with {RNA-As-Graphs}
  produces a large pool of candidate sequences with target topologies},}\
  }\href@noop {} {\bibfield  {journal} {\bibinfo  {journal} {J. Struct. Biol.}\
  }\textbf {\bibinfo {volume} {209}},\ \bibinfo {pages} {107438} (\bibinfo
  {year} {2020})}\BibitemShut {NoStop}%
\bibitem [{\citenamefont {Rolband}\ \emph {et~al.}(2022)\citenamefont
  {Rolband}, \citenamefont {Beasock}, \citenamefont {Wang}, \citenamefont
  {Shu}, \citenamefont {Dinman}, \citenamefont {Schlick}, \citenamefont {Zhou},
  \citenamefont {Kieft}, \citenamefont {Chen}, \citenamefont {Bussi} \emph
  {et~al.}}]{Rolband2022}%
  \BibitemOpen
  \bibfield  {author} {\bibinfo {author} {\bibfnamefont {L.}~\bibnamefont
  {Rolband}}, \bibinfo {author} {\bibfnamefont {D.}~\bibnamefont {Beasock}},
  \bibinfo {author} {\bibfnamefont {Y.}~\bibnamefont {Wang}}, \bibinfo {author}
  {\bibfnamefont {Y.-G.}\ \bibnamefont {Shu}}, \bibinfo {author} {\bibfnamefont
  {J.~D.}\ \bibnamefont {Dinman}}, \bibinfo {author} {\bibfnamefont
  {T.}~\bibnamefont {Schlick}}, \bibinfo {author} {\bibfnamefont
  {Y.}~\bibnamefont {Zhou}}, \bibinfo {author} {\bibfnamefont {J.~S.}\
  \bibnamefont {Kieft}}, \bibinfo {author} {\bibfnamefont {S.-J.}\ \bibnamefont
  {Chen}}, \bibinfo {author} {\bibfnamefont {G.}~\bibnamefont {Bussi}},  \emph
  {et~al.},\ }\bibfield  {title} {\enquote {\bibinfo {title} {Biomotors, viral
  assembly, and {RNA} nanobiotechnology: Current achievements and future
  directions},}\ }\href@noop {} {\bibfield  {journal} {\bibinfo  {journal}
  {Comput. Struct. Biotechnol. J.}\ } (\bibinfo {year} {2022})}\BibitemShut
  {NoStop}%
\bibitem [{\citenamefont {Herrero}\ \emph {et~al.}(2022)\citenamefont
  {Herrero}, \citenamefont {Stahl}, \citenamefont {Erbar}, \citenamefont
  {Maxeiner}, \citenamefont {Schlegel}, \citenamefont {Bacic}, \citenamefont
  {Cavalcanti}, \citenamefont {Schroer}, \citenamefont {Svergun}, \citenamefont
  {Sahin} \emph {et~al.}}]{Herrero2022}%
  \BibitemOpen
  \bibfield  {author} {\bibinfo {author} {\bibfnamefont {J.~M.}\ \bibnamefont
  {Herrero}}, \bibinfo {author} {\bibfnamefont {T.}~\bibnamefont {Stahl}},
  \bibinfo {author} {\bibfnamefont {S.}~\bibnamefont {Erbar}}, \bibinfo
  {author} {\bibfnamefont {K.}~\bibnamefont {Maxeiner}}, \bibinfo {author}
  {\bibfnamefont {A.}~\bibnamefont {Schlegel}}, \bibinfo {author}
  {\bibfnamefont {T.}~\bibnamefont {Bacic}}, \bibinfo {author} {\bibfnamefont
  {L.}~\bibnamefont {Cavalcanti}}, \bibinfo {author} {\bibfnamefont
  {M.}~\bibnamefont {Schroer}}, \bibinfo {author} {\bibfnamefont
  {D.}~\bibnamefont {Svergun}}, \bibinfo {author} {\bibfnamefont
  {U.}~\bibnamefont {Sahin}},  \emph {et~al.},\ }\bibfield  {title} {\enquote
  {\bibinfo {title} {Ultra-compacted single self-amplifying {RNA} molecules as
  quintessential vaccines},}\ }\href@noop {} {\bibfield  {journal} {\bibinfo
  {journal} {Research Square}\ } (\bibinfo {year} {2022})},\ \bibinfo {note}
  {\url{https://doi.org/10.21203/rs.3.rs-2142761/v1}}\BibitemShut {NoStop}%
\end{thebibliography}%


\begin{thebibliography}{6}%
\makeatletter
\providecommand \@ifxundefined [1]{%
 \@ifx{#1\undefined}
}%
\providecommand \@ifnum [1]{%
 \ifnum #1\expandafter \@firstoftwo
 \else \expandafter \@secondoftwo
 \fi
}%
\providecommand \@ifx [1]{%
 \ifx #1\expandafter \@firstoftwo
 \else \expandafter \@secondoftwo
 \fi
}%
\providecommand \natexlab [1]{#1}%
\providecommand \enquote  [1]{``#1''}%
\providecommand \bibnamefont  [1]{#1}%
\providecommand \bibfnamefont [1]{#1}%
\providecommand \citenamefont [1]{#1}%
\providecommand \href@noop [0]{\@secondoftwo}%
\providecommand \href [0]{\begingroup \@sanitize@url \@href}%
\providecommand \@href[1]{\@@startlink{#1}\@@href}%
\providecommand \@@href[1]{\endgroup#1\@@endlink}%
\providecommand \@sanitize@url [0]{\catcode `\\12\catcode `\$12\catcode
  `\&12\catcode `\#12\catcode `\^12\catcode `\_12\catcode `\%12\relax}%
\providecommand \@@startlink[1]{}%
\providecommand \@@endlink[0]{}%
\providecommand \url  [0]{\begingroup\@sanitize@url \@url }%
\providecommand \@url [1]{\endgroup\@href {#1}{\urlprefix }}%
\providecommand \urlprefix  [0]{URL }%
\providecommand \Eprint [0]{\href }%
\providecommand \doibase [0]{http://dx.doi.org/}%
\providecommand \selectlanguage [0]{\@gobble}%
\providecommand \bibinfo  [0]{\@secondoftwo}%
\providecommand \bibfield  [0]{\@secondoftwo}%
\providecommand \translation [1]{[#1]}%
\providecommand \BibitemOpen [0]{}%
\providecommand \bibitemStop [0]{}%
\providecommand \bibitemNoStop [0]{.\EOS\space}%
\providecommand \EOS [0]{\spacefactor3000\relax}%
\providecommand \BibitemShut  [1]{\csname bibitem#1\endcsname}%
\let\auto@bib@innerbib\@empty
\bibitem [{\citenamefont {Daoud}\ and\ \citenamefont
  {Joanny}(1981)}]{DaoudJoanny1981}%
  \BibitemOpen
  \bibfield  {author} {\bibinfo {author} {\bibfnamefont {M.}~\bibnamefont
  {Daoud}}\ and\ \bibinfo {author} {\bibfnamefont {J.~F.}\ \bibnamefont
  {Joanny}},\ }\href@noop {} {\bibfield  {journal} {\bibinfo  {journal} {J.
  Physique}\ }\textbf {\bibinfo {volume} {42}},\ \bibinfo {pages} {1359}
  (\bibinfo {year} {1981})}\BibitemShut {NoStop}%
\bibitem [{\citenamefont {Rosa}\ and\ \citenamefont
  {Everaers}(2017)}]{RosaEveraersPRE2017}%
  \BibitemOpen
  \bibfield  {author} {\bibinfo {author} {\bibfnamefont {A.}~\bibnamefont
  {Rosa}}\ and\ \bibinfo {author} {\bibfnamefont {R.}~\bibnamefont
  {Everaers}},\ }\href@noop {} {\bibfield  {journal} {\bibinfo  {journal}
  {Phys. Rev. E}\ }\textbf {\bibinfo {volume} {95}},\ \bibinfo {pages} {012117}
  (\bibinfo {year} {2017})}\BibitemShut {NoStop}%
\bibitem [{\citenamefont {Vaupoti\v{c}}\ \emph {et~al.}(2022)\citenamefont
  {Vaupoti\v{c}}, \citenamefont {Rosa}, \citenamefont {Podgornik},
  \citenamefont {Tubiana},\ and\ \citenamefont
  {{Bo\v{z}i\v{c}}}}]{Vaupotic2022}%
  \BibitemOpen
  \bibfield  {author} {\bibinfo {author} {\bibfnamefont {D.}~\bibnamefont
  {Vaupoti\v{c}}}, \bibinfo {author} {\bibfnamefont {A.}~\bibnamefont {Rosa}},
  \bibinfo {author} {\bibfnamefont {R.}~\bibnamefont {Podgornik}}, \bibinfo
  {author} {\bibfnamefont {L.}~\bibnamefont {Tubiana}}, \ and\ \bibinfo
  {author} {\bibfnamefont {A.}~\bibnamefont {{Bo\v{z}i\v{c}}}},\ }\href@noop {}
  {\enquote {\bibinfo {title} {Viral {RNA} as a branched polymer},}\ }
  (\bibinfo {year} {2022}),\ \Eprint {http://arxiv.org/abs/arXiv:2212.00829}
  {arXiv:2212.00829 [physics.bio-ph]} \BibitemShut {NoStop}%
\bibitem [{\citenamefont {Lorenz}\ \emph {et~al.}(2011)\citenamefont {Lorenz},
  \citenamefont {Bernhart}, \citenamefont {H{\"o}ner~zu Siederdissen},
  \citenamefont {Tafer}, \citenamefont {Flamm}, \citenamefont {Stadler},\ and\
  \citenamefont {Hofacker}}]{Lorenz2011}%
  \BibitemOpen
  \bibfield  {author} {\bibinfo {author} {\bibfnamefont {R.}~\bibnamefont
  {Lorenz}}, \bibinfo {author} {\bibfnamefont {S.~H.}\ \bibnamefont
  {Bernhart}}, \bibinfo {author} {\bibfnamefont {C.}~\bibnamefont {H{\"o}ner~zu
  Siederdissen}}, \bibinfo {author} {\bibfnamefont {H.}~\bibnamefont {Tafer}},
  \bibinfo {author} {\bibfnamefont {C.}~\bibnamefont {Flamm}}, \bibinfo
  {author} {\bibfnamefont {P.~F.}\ \bibnamefont {Stadler}}, \ and\ \bibinfo
  {author} {\bibfnamefont {I.~L.}\ \bibnamefont {Hofacker}},\ }\href@noop {}
  {\bibfield  {journal} {\bibinfo  {journal} {Algorithms Mol. Biol.}\ }\textbf
  {\bibinfo {volume} {6}},\ \bibinfo {pages} {1} (\bibinfo {year}
  {2011})}\BibitemShut {NoStop}%
\bibitem [{\citenamefont {Singaram}\ \emph {et~al.}(2016)\citenamefont
  {Singaram}, \citenamefont {Gopal},\ and\ \citenamefont
  {Ben-Shaul}}]{Singaram2016}%
  \BibitemOpen
  \bibfield  {author} {\bibinfo {author} {\bibfnamefont {S.~W.}\ \bibnamefont
  {Singaram}}, \bibinfo {author} {\bibfnamefont {A.}~\bibnamefont {Gopal}}, \
  and\ \bibinfo {author} {\bibfnamefont {A.}~\bibnamefont {Ben-Shaul}},\
  }\href@noop {} {\bibfield  {journal} {\bibinfo  {journal} {J. Phys. Chem. B}\
  }\textbf {\bibinfo {volume} {120}},\ \bibinfo {pages} {6231} (\bibinfo {year}
  {2016})}\BibitemShut {NoStop}%
\bibitem [{\citenamefont {Parisse}\ and\ \citenamefont {{De
  Graeve}}(2022)}]{Giac}%
  \BibitemOpen
  \bibfield  {author} {\bibinfo {author} {\bibfnamefont {B.}~\bibnamefont
  {Parisse}}\ and\ \bibinfo {author} {\bibfnamefont {R.}~\bibnamefont {{De
  Graeve}}},\ }\href
  {https://www-fourier.univ-grenoble-alpes.fr/~parisse/giac.html} {\enquote
  {\bibinfo {title} {Giac/xcas, v1.9.0},}\ } (\bibinfo {year} {2022}),\
  \bibinfo {note}
  {\url{https://www-fourier.univ-grenoble-alpes.fr/~parisse/giac.html}}\BibitemShut
  {NoStop}%
\end{thebibliography}%

\end{document}


\title{SUPPLEMENTARY MATERIAL:\\ Scaling properties of RNA as a randomly branching polymer}

\author{Domen Vaupoti\v{c}}
\affiliation{Department of Theoretical Physics, Jo\v{z}ef Stefan Institute, Jamova 39, 1000 Ljubljana, Slovenia}
\author{Angelo Rosa}
\affiliation{Scuola Internazionale Superiore di Studi Avanzati (SISSA), Via Bonomea 265, 34136 Trieste, Italy}
\author{Luca Tubiana}
\affiliation{Department of Physics, University of Trento, via Sommarive 14, 38123 Trento, Italy}
\affiliation{INFN-TIFPA, Trento Institute for Fundamental Physics and Applications, via Sommarive 14, 38123 Trento, Italy}
\author{An\v{z}e Bo\v{z}i\v{c}}
\affiliation{Department of Theoretical Physics, Jo\v{z}ef Stefan Institute, Jamova 39, 1000 Ljubljana, Slovenia}
\email{anze.bozic@ijs.si}

\date{\today}

\maketitle

\newpage

\section*{Justification of the ansatz in Equation~(13) in the main text}
Equation~(11) in the main text is exact only in the case of {\em ideal} branching polymers, i.e., polymers with no monomer-monomer interactions for which the only contribution to the free energy of the system comes from branching. For such a situation, \citet{DaoudJoanny1981} have shown that the partition function of branching polymers with size $N$ is given by:
\begin{equation}
\label{eq:DaoudJoannyZ}
\Z_N^{\rm ideal} = \frac{I_1(2\lambda N)}{\lambda N} ,
\end{equation}
where $I_1(x)$ is the modified Bessel function of the first kind and $\lambda$ is the branching fugacity (related to the branching probability per node). From the known asymptotic behavior for Bessel functions, we can write down the large-$N$ limit of $\Z_N^{\rm ideal}$ as
\begin{equation}
\label{eq:DaoudJoannyZ-LargeN}
\lim_{\lambda N\rightarrow\infty} \Z_N^{\rm ideal} \simeq \frac{e^{2\lambda N}}{2\sqrt{\pi} \, (\lambda N)^{3/2}}.
\end{equation}
Even though Eq.~(11) in the main text holds only in the ideal case, it was shown by \citet{RosaEveraersPRE2017} that it remains remarkably accurate even for interacting branching polymers and hence should hold for RNA molecules as well.
In order to take advantage of this and employ Eq.~(11) to extract the exponent $\varepsilon$ from the branch weight distribution functions of RNA molecules, we can introduce the following ansatz [i.e., Eq.~(13) in the main text] for the partition function of interacting polymers:
\begin{equation}
\label{eq:ansatz}
\Z_N = \frac{I_\beta(2\lambda N)}{\left(\lambda N \right)^\beta}.
\end{equation}
In a similar fashion as for Eq.~\eqref{eq:DaoudJoannyZ-LargeN}, we can also determine the large-$N$ limit of Eq.~\eqref{eq:ansatz}:
\begin{equation}
\label{eq:ansatz-LargeN}
\lim_{\lambda N\rightarrow\infty} \Z_N \simeq \frac{e^{2\lambda N}}{2\sqrt{\pi} \, (\lambda N)^{\beta+1/2}} .
\end{equation}
By using the asymptotic expansion in Eq.~\eqref{eq:ansatz-LargeN}, we can show that the distribution of branch weights $p(\nbr)$ given by Eq.~(11) in the main text follows the power-law-like behaviour in the mid-weight regime, $1/(2\lambda) \ll \nbr \leqslant N/2$: 
\begin{align}
\label{eq:p-sup}
\nonumber p(\nbr) &= \frac{\Z_\nbr \Z_{N-\nbr-1}}{\sum_{\nbr=0}^{N-1} \Z_\nbr \Z_{N-\nbr-1}} \propto \Z_\nbr \Z_{N-\nbr-1} \\ 
\nonumber & \sim e^{2\lambda \nbr} e^{2\nbr\left(N-\nbr-1\right)} \left(\lambda^2 \nbr \left(N-\nbr-1\right) \right)^{-(\beta+1/2)} \\
&= \frac{e^{2\lambda(N-1)}}{\lambda^{-2(\beta+1/2)}}\left(\nbr\left(N-\nbr-1\right)\right)^{-(\beta+1/2)} \propto \left(\nbr\left(N-\nbr-1\right)\right)^{-(\beta+1/2)} .
\end{align}
By comparing Eq.~\eqref{eq:p-sup} to Eq.~(12) in the main text, it follows that $\beta=3/2-\varepsilon$.

\newpage

\section*{RNA sequence length and the size of the tree representation of its structure}

\begin{figure*}[!htp]
  \centering
  \includegraphics[width=0.705\linewidth]{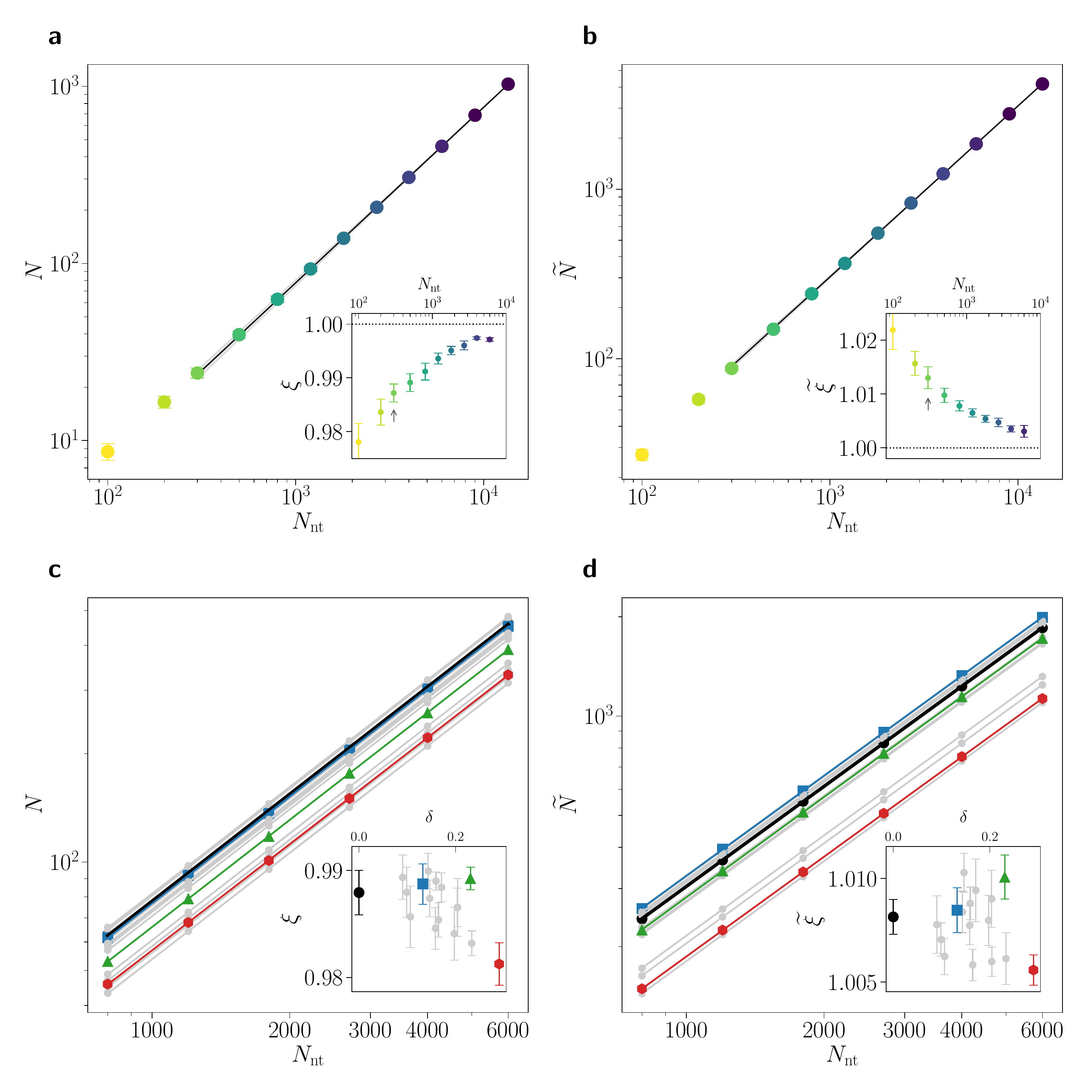}
  \caption{Scaling of the number of {\bf (a)} tree edges $N$ and {\bf (b)} expanded tree edges $\widetilde N$ with RNA sequence length $N_{\rm nt}$ of a random RNA with uniform nucleotide composition. The relationships between these quantities, $N\sim N_\mathrm{nt}^\xi$ and $\widetilde{N}\sim N_\mathrm{nt}^{\widetilde{\xi}}$, are almost completely linear ($\xi=\widetilde{\xi}\approx 1$, insets of panels a and b). \rev{Scaling of the number of {\bf (c)} tree edges $N$ and {\bf (d)} expanded tree edges $\widetilde N$ with RNA sequence length $N_{\rm nt}$ for random RNA sequences of $16$ different nucleotide compositions (listed in Table~\ref{tab:nuccomp}). Insets of panels c and d show that the linear relationships persist even with varying nucleotide composition.}}
\label{fig:S1}
\end{figure*}

By mapping an RNA secondary structure of length $N_\mathrm{nt}$ to a tree, one obtains a planar tree graph with $N+1$ nodes and $N$ weighted edges (see Sec.~II in the main text). The RNA tree can be furthermore expanded in such a way that all edges have unit length, resulting in an expanded tree with $\widetilde N$ edges \rev{(with $\widetilde N$ in turn corresponding to the total number of base pairs). Given an average stem length ${b}$ and number of unpaired nucleotides $N_\mathrm{unpaired}$, these quantities are related through
\begin{equation}
N_\mathrm{nt}=2{b}N+N_\mathrm{unpaired}=2\widetilde{N}+N_\mathrm{unpaired},
\end{equation}
since we have $\widetilde{N}={b}N$. The average base-pairing probability and the average stem length ${b}$ are both independent of the length of the RNA in the limit of long RNAs, meaning that the ratio $N_\mathrm{unpaired}/N_\mathrm{nt}$ is not a function of sequence length. Furthermore, while changing the composition of the RNA sequence influences the amount of base pairs in its secondary structure, the percentage of base pairs (or, alternatively, average stem length ${b}$) remains independent of the length of the sequence. Therefore, we can expect that the RNA sequence length $N_\mathrm{nt}$ and the number of edges in both of its tree representations ($N$ and $\widetilde{N}$) will be linearly correlated and we can use them interchangeably. This is indeed so, as Fig.~\ref{fig:S1} confirms: both $N$ and $\widetilde{N}$ are almost perfectly linearly correlated with $N_\mathrm{nt}$, the more so the larger the RNA tree. Changing the sequence composition of the RNA also does not compromise this linear correlation, but only changes the prefactor of the scaling.}

\clearpage
\newpage

\section*{Scaling exponents and nucleotide composition of RNA}

Most of our work focuses on uniformly random RNA sequences where the frequency of each nucleotide is identical, $f(b)=0.25\;\forall b$. To verify that the conclusions we obtain regarding their scaling exponents are also valid in general for an arbitrary nucleotide composition, we selected $16$ different nucleotide compositions (Table~\ref{tab:nuccomp}) and evaluated their distance from the uniform random composition by a simple Euclidean metric,
\begin{equation}
\delta^2=\sum_{b\in\{\A,\C,\G,\U\}}\left(f(b)-\frac14\right)^2,
\end{equation}
which correlates well with improved statistical measures such as Jensen-Shannon divergence. The $16$ different nucleotide compositions cover a wide array of possibilities, as they were chosen in such a way that they cover the convex hull of the space of $\sim 1800$ viral genomes of different positive single-stranded RNA viruses and as such represent the most extreme cases within this dataset.

\begin{figure*}[!htp]
  \centering
  \includegraphics[width=0.7\linewidth]{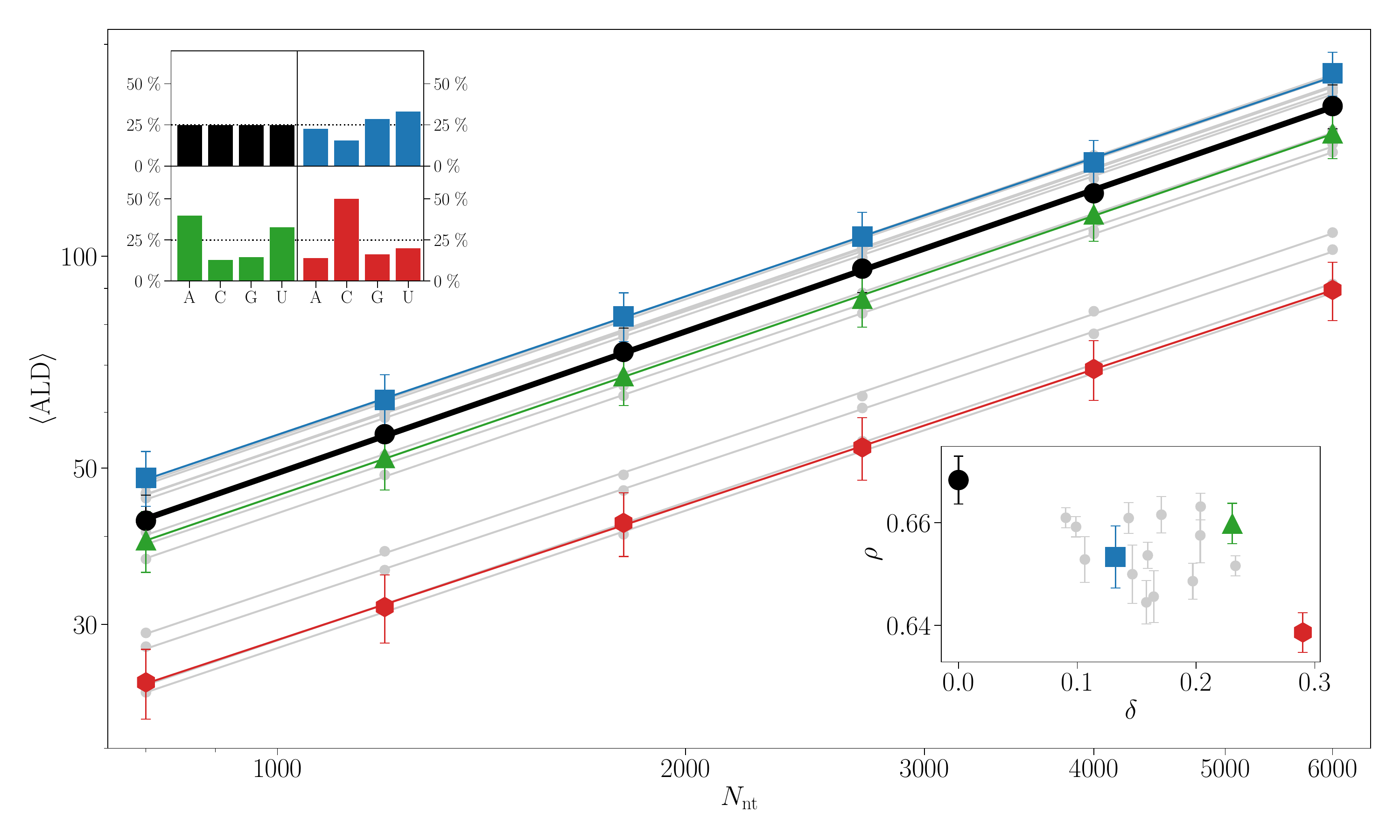}
  \caption{Scaling of $\ALD$ with RNA sequence length $N_\mathrm{nt}$ for 16 different nucleotide compositions of random RNA sequences (Table~\ref{tab:nuccomp}). Each datapoint (for a given nucleotide composition and sequence length) contains 200 random sequences. Smaller insets show the nucleotide composition for four different examples. Larger inset shows the scaling exponent $\rho$ obtained for random RNAs with different nucleotide composition, as measured by Euclidean distance $\delta$ from uniformly random composition.}
\label{fig:S4}
\end{figure*}

\begin{figure*}[!htp]
  \centering
  \includegraphics[width=0.7\linewidth]{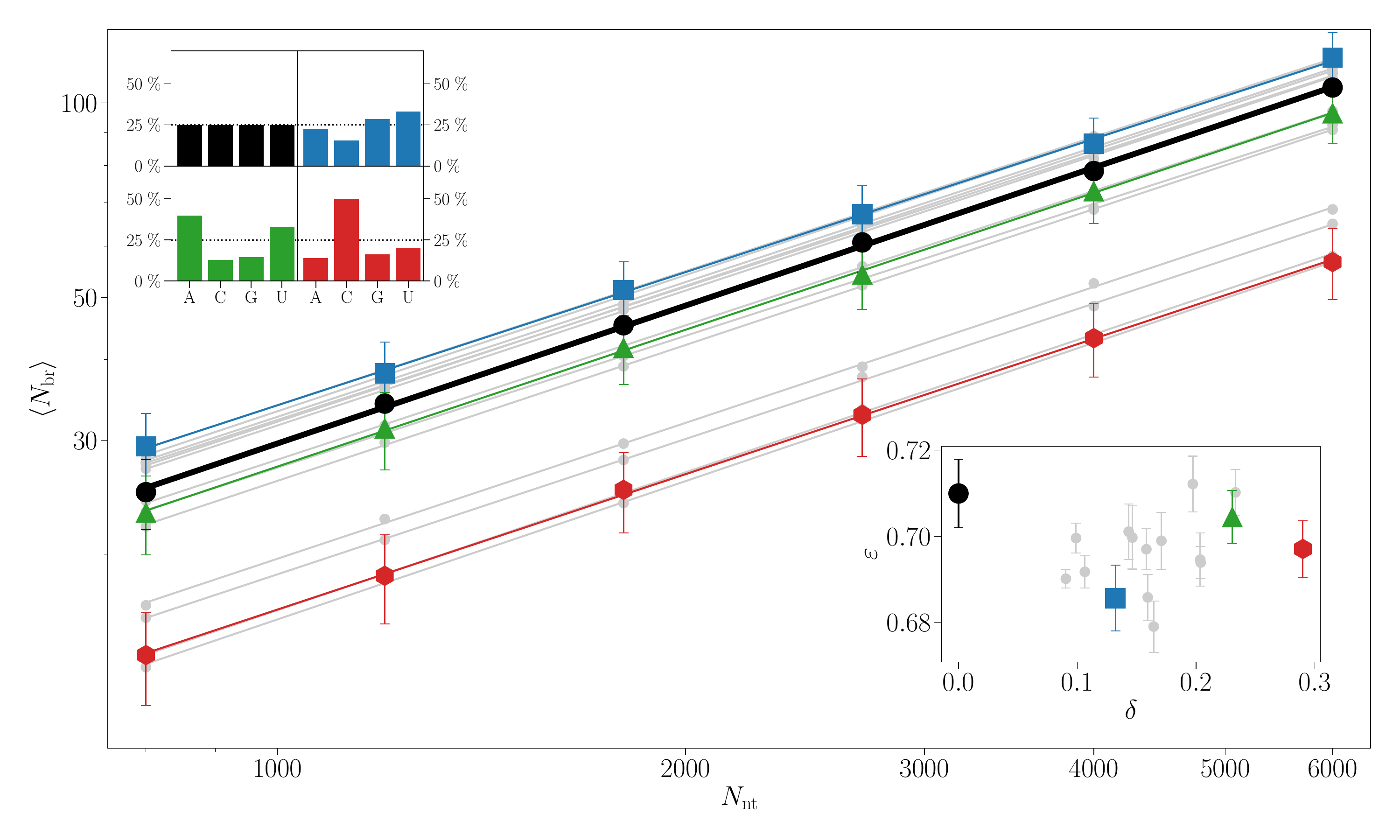}
  \caption{Scaling of $\Nbr$ with RNA sequence length $N_\mathrm{nt}$ for 16 different nucleotide compositions of random RNA sequences (Table~\ref{tab:nuccomp}). Each datapoint (for a given nucleotide composition and sequence length) contains 200 random sequences. Smaller insets show the nucleotide composition for four different examples. Larger inset shows the scaling exponent $\varepsilon$ obtained for random RNAs with different nucleotide composition, as measured by Euclidean distance $\delta$ from uniformly random composition.}
\label{fig:S5}
\end{figure*}

Both the scaling of the $\ALD$ (Fig.~\ref{fig:S4}) and $\Nbr$ (Fig.~\ref{fig:S5}) with sequence length for random RNAs with different nucleotide composition show only a small variation in the scaling exponents $\rho$ and $\varepsilon$. This holds true even when the nucleotide composition differs significantly from that of a uniformly random RNA. The observed differences in scaling are thus solely a consequence of the prefactor of the scaling law, which has previously already been connected to the differences in the percentage of base pairs in the RNA structures at different compositions~\cite{Vaupotic2022}.

\begin{table}[!htp]
\caption{Nucleotide composition of $16$ different sets of RNA used for the comparison of the scaling behaviour of random RNA sequences with different nucleotide compositions. }
\renewcommand{\arraystretch}{1.4} 
\begin{tabular}{cccccc}
\hline\hline
Composition & $f(\A)$ & $f(\C)$ & $f(\G)$ & $f(\U)$ & $\delta$\\
\hline
uniform & $0.2500$ & $0.2500$ & $0.2500$ & $0.2500$ & $0.0000$ \\
1 & $0.1789$ & $0.2639$ & $0.3048$ & $0.2524$ & $0.0909$ \\
2 & $0.3298$ & $0.2336$ & $0.2427$ & $0.1939$ & $0.0992$ \\
3 & $0.2109$ & $0.2469$ & $0.2042$ & $0.3379$ & $0.1066$ \\
4 & $0.2255$ & $0.1560$ & $0.2864$ & $0.3321$ & $0.1322$ \\
5 & $0.2095$ & $0.1630$ & $0.2735$ & $0.3539$ & $0.1433$ \\
6 & $0.3366$ & $0.2920$ & $0.1409$ & $0.2304$ & $0.1468$ \\
7 & $0.2249$ & $0.3595$ & $0.1408$ & $0.2748$ & $0.1587$ \\
8 & $0.2202$ & $0.1624$ & $0.2382$ & $0.3792$ & $0.1594$ \\
9 & $0.1550$ & $0.3820$ & $0.2352$ & $0.2277$ & $0.1648$ \\
10 & $0.2505$ & $0.1427$ & $0.2259$ & $0.3809$ & $0.1710$ \\
11 & $0.2252$ & $0.4057$ & $0.1311$ & $0.2380$ & $0.1978$ \\
12 & $0.2784$ & $0.1302$ & $0.1904$ & $0.4010$ & $0.2038$ \\
13 & $0.1494$ & $0.3877$ & $0.3082$ & $0.1547$ & $0.2039$ \\
14 & $0.3973$ & $0.1297$ & $0.1452$ & $0.3277$ & $0.2306$ \\
15 & $0.1877$ & $0.4480$ & $0.1486$ & $0.2157$ & $0.2336$ \\
16 & $0.1392$ & $0.4989$ & $0.1635$ & $0.1984$ & $0.2905$ \\ 
\hline\hline
\end{tabular}
\label{tab:nuccomp}
\end{table}

\clearpage
\newpage

\section*{Scaling exponents and multiloop energy parameters of the folding software}

In the commonly used linear model for the energy of multiloop formation,
\begin{equation}
E_\mathrm{multiloop}=E_0+E_\mathrm{br}\times[\mathrm{branches}]+E_\mathrm{un}\times[\textrm{unpaired nucleotides}],
\end{equation}
different versions of ViennaRNA software~\cite{Lorenz2011} (specifically, those before and after v2.0) use significantly different values of the multiloop energy parameters. In particular, the largest difference lies in the energy of branch formation $E_\mathrm{br}$ which changes sign in the two parameter sets, either inhibiting large multiloop formations in the older versions of the software (positive value of the parameter) or promoting them in the newer versions (negative value of the parameter; see Table~\ref{tab:multiloop}). To consistently compare the influence of the multiloop energy parameters on the scaling behaviour of RNA structures, we use the energy parameters from ViennaRNA v2.4 and use either its version of multiloop energy parameters or we modify {\em solely} the multiloop energy parameters with those of older versions, resulting in an energy parameter set we term ViennaRNA-mod. All other parameters in the ViennaRNA-mod set are left unchanged, i.e., equal to the set used in ViennaRNA v2.4. In this way, we can be sure that any differences in scaling we might observe are due to the changes in the multiloop energies. The values for the two sets of multiloop energy parameters used are given in Table~\ref{tab:multiloop}.

\begin{figure*}[!htp]
  \centering
  \includegraphics[width=\linewidth]{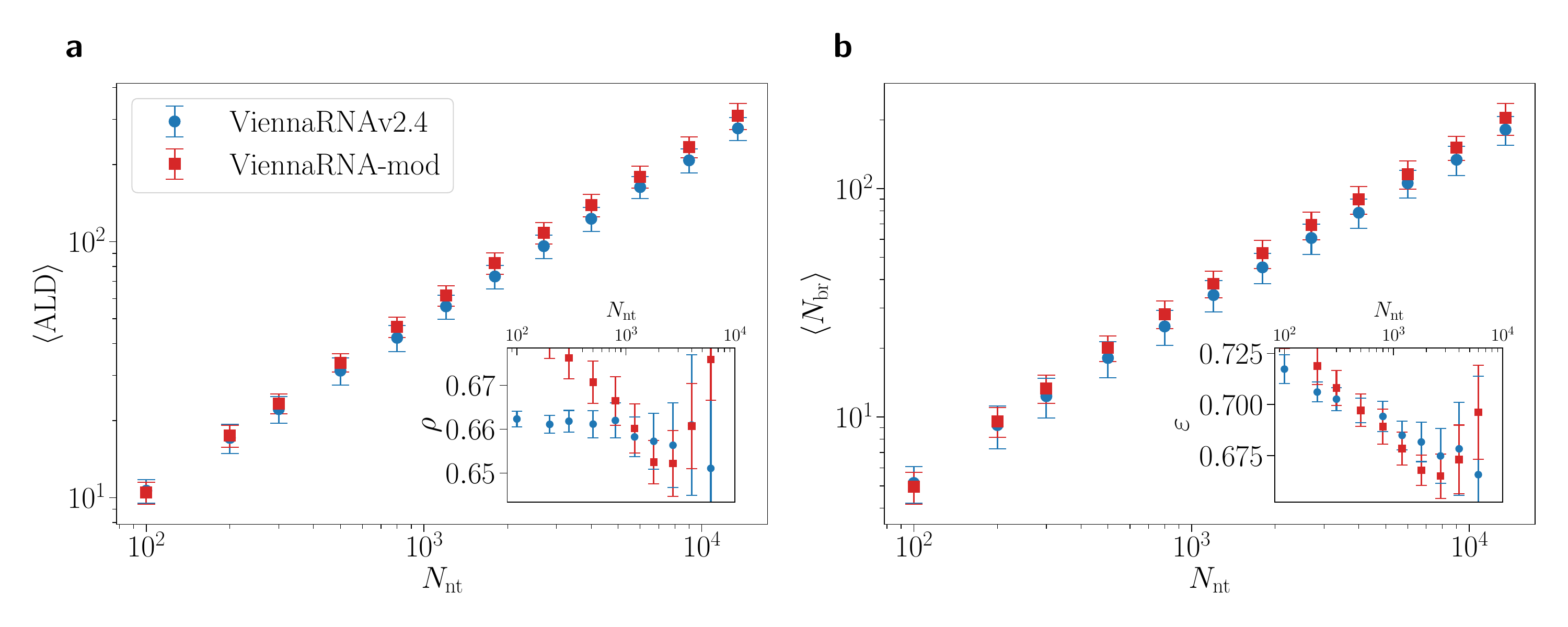}
  \caption{Scaling of {\bf (a)} $\ALD$ and {\bf (b)} $\Nbr$ with RNA sequence length $N_{\rm nt}$ using two different multiloop energy parameters, based either on the current version of ViennaRNA (ViennaRNAv2.4) or on its older versions ($<2.0$, ViennaRNA-mod). The insets show how the scaling exponents change as we progressively narrow the fit region to longer sequence lengths.}
\label{fig:s2}
\end{figure*}

Figure~\ref{fig:s2} shows the scaling of both $\ALD$ and $\Nbr$ with sequence length for uniformly random RNA folded using the two different multiloop energy parameters. Despite the fact that the two sets of parameters tend to result in quite different node degree distributions of the resulting RNA structures (see Ref.~\cite{Vaupotic2022} for details), we can see that both scaling exponents $\rho$ and $\varepsilon$ seem to converge to the same values.

\begin{table}[!htp]
\caption{Multiloop energy parameters used in the current version of ViennaRNA (v2.4) and those from the older versions of the software ($<2.0$), which we use in a modified set of energy parameters termed ViennaRNA-mod.}
\renewcommand{\arraystretch}{1.4} 
\begin{tabular}{lccc}
\hline\hline
Multiloop parameters & $\phantom{+}E_0$ & $\phantom{+}E_\mathrm{un}$ & $\phantom{+}E_\mathrm{br}$\\
\hline
ViennaRNAv2.4 & $\phantom{0}9.3$ & $\phantom{+}0.0$ & $-0.9$ \\
ViennaRNA-mod & $\phantom{0}3.4$ & $\phantom{+}0.0$ & $\phantom{+}0.4$\\
\hline\hline
\end{tabular}
\label{tab:multiloop}
\end{table}

\clearpage
\newpage

\section*{Scaling exponents and Pr\"ufer-shuffled RNA trees}

\begin{figure}[!htp]
\centering
\includegraphics[width=0.75\linewidth]{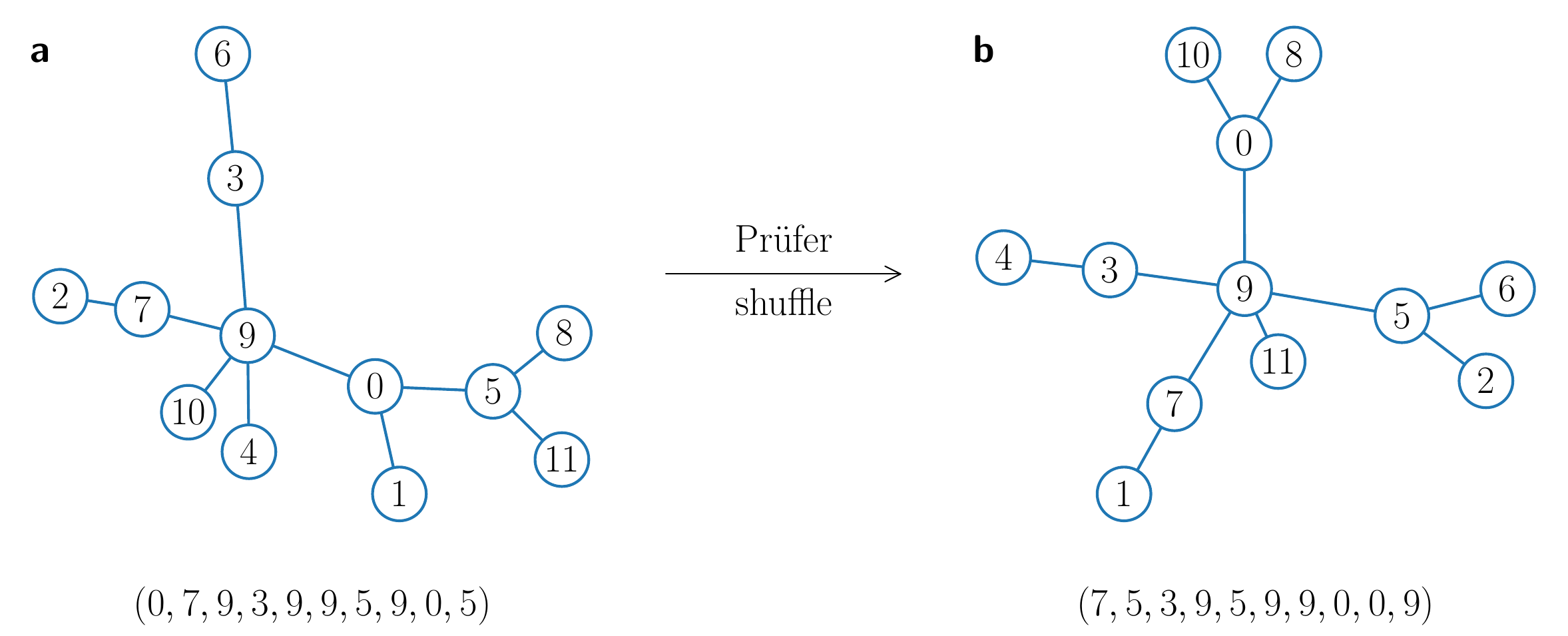}
\caption{{\bf (a)} Example of a Pr\" ufer sequence of a small labelled tree with $N=11$ edges and {\bf (b)} Pr\" ufer-shuffled version of the same tree (obtained by shuffling the original Pr\" ufer sequence to obtain the Pr\" ufer sequence of the shuffled tree). Note that while the original and the shuffled tree are different, they have the same distribution of node degrees (e.g., one node of degree $5$ and two nodes of degree $3$).}
\label{fig:prufer-example}
\end{figure}

Every labelled planar tree graph with $N$ edges can be uniquely represented by its Pr\" ufer sequence of length $N-1$. \rev{The Pr\" ufer sequence is generated iteratively from a labelled tree by successively removing the peripheral node (leaf) with the smallest label, and adding the number of the node to which it was connected as the next element in the sequence. The details of this procedure are described in Ref.~\cite{Singaram2016}. An example of a small tree together with its Pr\" ufer sequence is shown in Fig.~\ref{fig:prufer-example}a.}

\begin{figure}[!htp]
\centering
\includegraphics[width=0.7\linewidth]{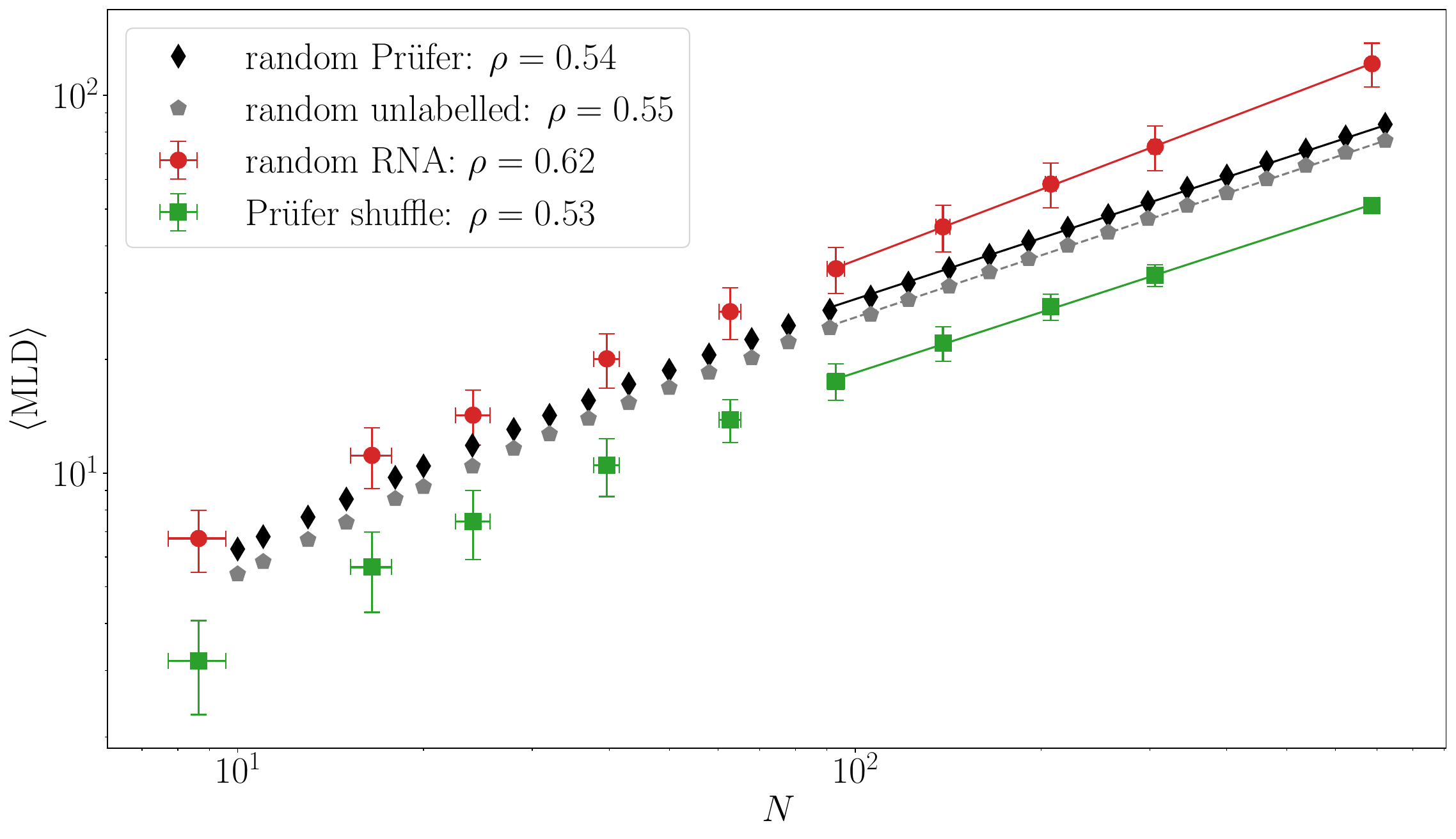}
\caption{Scaling of $\langle\mathrm{MLD}\rangle$ with the number of tree edges $N$ for uniformly random RNA sequences of different lengths (red circles) and their Pr\"ufer-shuffled versions (green squares). The error bars shows standard deviations of averaging over $500$ permutations of the corresponding Pr\"ufer sequence of the original RNA tree. Black diamonds show the scaling of random Pr\"ufer sequences, corresponding to a sampling of random labelled trees, and gray pentagons show the scaling behaviour of random unlabelled (non-isomorphic) trees. Lines show power law fits of the form $\langle\mathrm{MLD}\rangle=\alpha N^\rho$.}
\label{fig:prufer}
\end{figure}

\begin{figure}[!htp]
\centering
\includegraphics[width=0.7\linewidth]{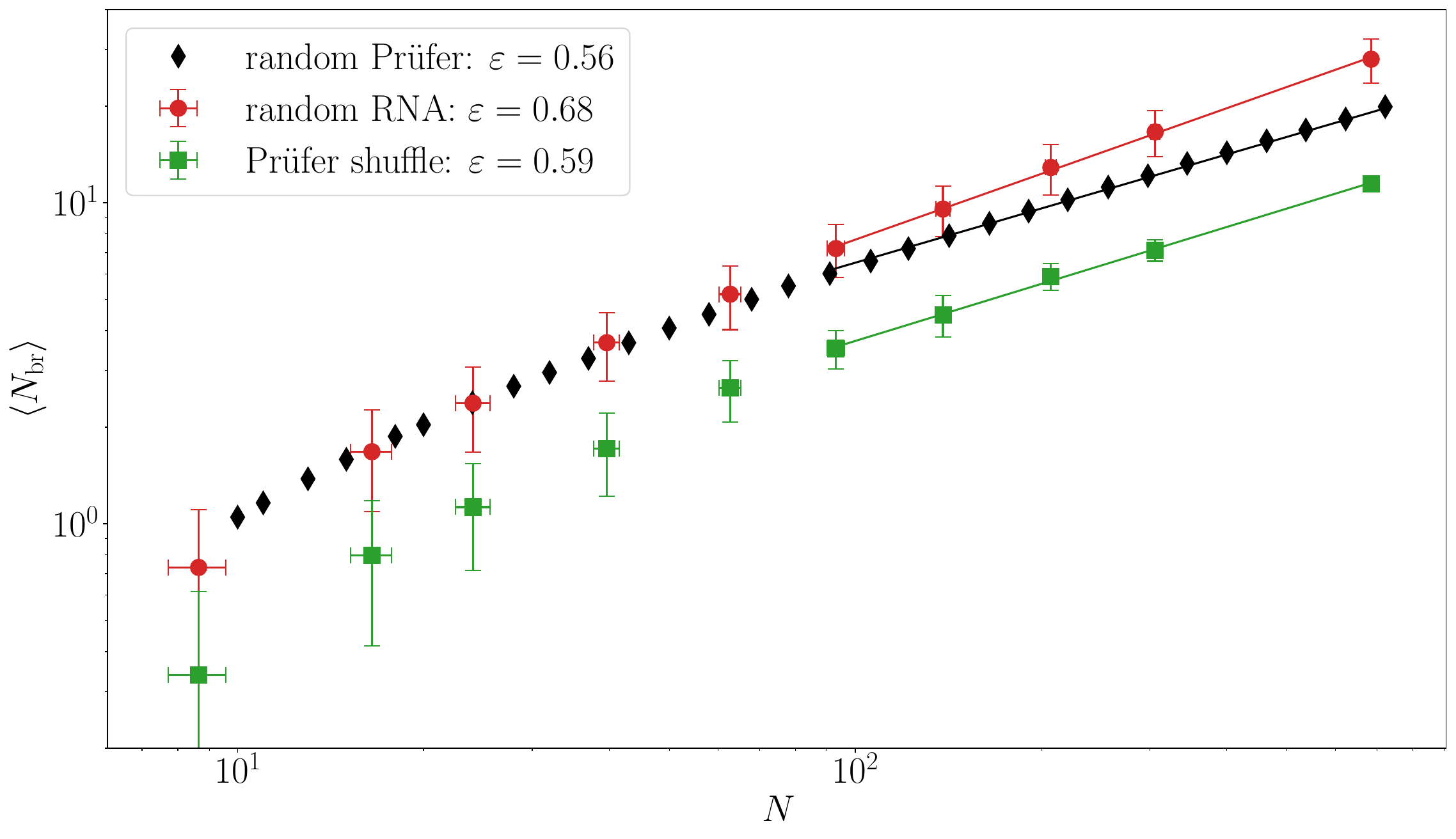}
\caption{Scaling of $\langle N_\mathrm{br}\rangle$ with the number of tree edges $N$ for uniformly random RNA sequences of different lengths (red circles) and their Pr\"ufer-shuffled versions (green squares). Each square represents an average over $500$ permutations of the corresponding Pr\"ufer sequence of the original RNA tree. Black diamonds show the scaling of random Pr\"ufer sequences, corresponding to a sampling of random labelled trees. Lines show power law fits of the form $\langle N_\mathrm{br}\rangle=\alpha N^\varepsilon$.}
\label{fig:prufer2}
\end{figure}

Permutation of the Pr\" ufer sequence of a tree (Pr\" ufer shuffle) results in a different tree with an identical {\em distribution of node degrees} (see Fig.~\ref{fig:prufer-example} for an example). Thus, by permuting the Pr\" ufer sequences of RNA structures mapped to trees, we can obtain different trees whose node degree distribution exactly matches that of the original RNA and can be considered RNA-like. Figures~\ref{fig:prufer} and~\ref{fig:prufer2} show the scaling of $\MLD$ and $\Nbr$ with the number of edges $N$ for random RNA and its Pr\" ufer-shuffled versions (each sequence averaged over $500$ Pr\" ufer shuffles). For comparison, we also show the scaling of randomly generated Pr\" ufer sequences, corresponding to random labelled trees with $N$ edges.

We also note that while RNA trees can be considered unlabelled, Pr\"ufer sequence representation in general operates in the space of labelled trees and is therefore sensitive to node labelling. To make sure that we can nonetheless compare the scaling of RNA trees and their Pr\" ufer-shuffled versions, we also generated random unlabelled (non-isomorphic) trees with $N$ edges as implemented in Giac/Xcas software~\cite{Giac}. The scaling of $\MLD$ for both random labelled and unlabelled trees is essentially the same (Fig.~\ref{fig:prufer}), which justifies {\em post hoc} our initial comparison of RNA trees and their Pr\" ufer-shuffled versions.

\clearpage
\newpage

\section*{Scaling exponents $t$ and $\theta$ of the {RdC} distribution of path lengths}

\begin{figure*}[!htp]
  \centering
  \includegraphics[width=0.9\linewidth]{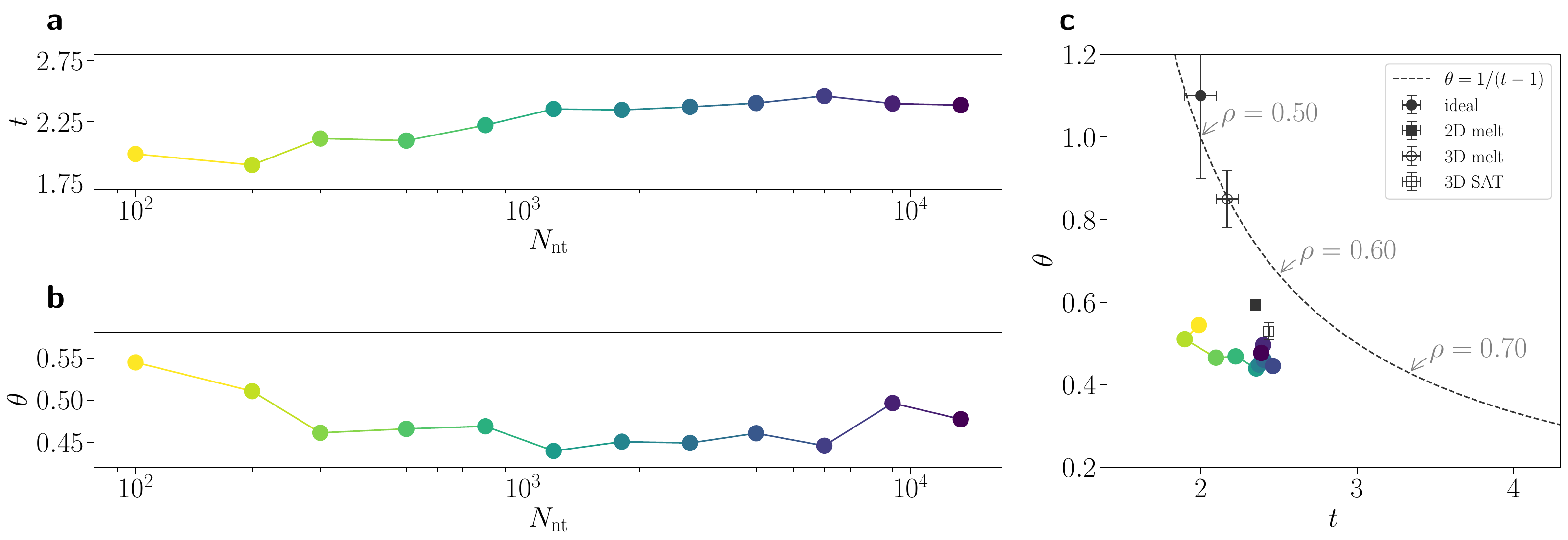}
  \caption{Parameters {\bf (a)} $t$ and {\bf (b)} $\theta$ of the RdC distribution obtained from path length distributions of uniformly random RNA sequences of different lengths. Each point represents a fit to the distribution of path lengths of $200$ random sequences with $500$ structures for each. Panel {\bf (c)} shows the changes in the fitted exponents in the $(t,\theta)$ plane, where one can observe how the relationship between them changes with increasing sequence length (colour-coded in panels a and b). Also shown are the values for four different types of branched polymers, taken from Ref.~\cite{RosaEveraersPRE2017}. Dashed line in panel c shows the theoretical relationship between the two fit exponents, Eq.~\eqref{eq:relationship}.}
\label{fig:s3}
\end{figure*}

To obtain the scaling exponent $\rho$ from the distribution of path lengths $p(\ell)$ of individual RNA structures (or, more precisely, sets of RNA structures {\em at fixed length}), we use the two-parametric RdC distribution [Eq.~(6) in the main text] and fit it to the path length distribution. Typically, we do this for a set of random sequences and their structures at a fixed length and obtain the two fit parameters $t$ and $\theta$ (panels a and b of Fig.~\ref{fig:s3}). One can show that both fit parameters can be connected to the scaling exponent $\rho$---in this way, we can use the two parameters to obtain two independent estimates of the scaling exponent, $\rho_t$ and $\rho_\theta$, respectively. Under certain assumptions (see the discussion in Sec.~IV C in the main text), these two parameters are not independent but are related through
\begin{equation}
\label{eq:relationship}
    \theta=\frac{1}{t-1}.
\end{equation}
Figure~\ref{fig:s3}c shows how the RdC fit exponents for uniformly random RNAs of different lengths compare to Eq.~\eqref{eq:relationship}. For comparison, we also show the values of the RdC fit exponents for four different types of randomly branched polymers, taken from Ref.~\cite{RosaEveraersPRE2017}.

\FloatBarrier
\newpage

\bibliographystyle{apsrev4-1}
\bibliography{references}